\PassOptionsToPackage{unicode}{hyperref}
\PassOptionsToPackage{hyphens}{url}
\PassOptionsToPackage{dvipsnames,svgnames,x11names}{xcolor}
\documentclass[
  12pt]{article}

\usepackage{amsmath,amssymb}
\usepackage{iftex}
\ifPDFTeX
  \usepackage[T1]{fontenc}
  \usepackage[utf8]{inputenc}
  \usepackage{textcomp} 
\else 
  \usepackage{unicode-math}
  \defaultfontfeatures{Scale=MatchLowercase}
  \defaultfontfeatures[\rmfamily]{Ligatures=TeX,Scale=1}
\fi
\usepackage{lmodern}
\ifPDFTeX\else  
\fi
\IfFileExists{upquote.sty}{\usepackage{upquote}}{}
\IfFileExists{microtype.sty}{
  \usepackage[]{microtype}
  \UseMicrotypeSet[protrusion]{basicmath} 
}{}
\makeatletter
\@ifundefined{KOMAClassName}{
  \IfFileExists{parskip.sty}{%
    \usepackage{parskip}
  }{
    \setlength{\parindent}{0pt}
    \setlength{\parskip}{6pt plus 2pt minus 1pt}}
}{
  \KOMAoptions{parskip=half}}
\makeatother
\usepackage{xcolor}
\makeatletter
\ifx\paragraph\undefined\else
  \let\oldparagraph\paragraph
  \renewcommand{\paragraph}{
    \@ifstar
      \xxxParagraphStar
      \xxxParagraphNoStar
  }
  \newcommand{\xxxParagraphStar}[1]{\oldparagraph*{#1}\mbox{}}
  \newcommand{\xxxParagraphNoStar}[1]{\oldparagraph{#1}\mbox{}}
\fi
\ifx\subparagraph\undefined\else
  \let\oldsubparagraph\subparagraph
  \renewcommand{\subparagraph}{
    \@ifstar
      \xxxSubParagraphStar
      \xxxSubParagraphNoStar
  }
  \newcommand{\xxxSubParagraphStar}[1]{\oldsubparagraph*{#1}\mbox{}}
  \newcommand{\xxxSubParagraphNoStar}[1]{\oldsubparagraph{#1}\mbox{}}
\fi
\makeatother

\usepackage{longtable,booktabs,array}
\usepackage{calc} 
\usepackage{etoolbox}
\makeatletter
\patchcmd\longtable{\par}{\if@noskipsec\mbox{}\fi\par}{}{}
\makeatother
\IfFileExists{footnotehyper.sty}{\usepackage{footnotehyper}}{\usepackage{footnote}}
\makesavenoteenv{longtable}
\usepackage{graphicx}
\makeatletter
\def\maxwidth{\ifdim\Gin@nat@width>\linewidth\linewidth\else\Gin@nat@width\fi}
\def\maxheight{\ifdim\Gin@nat@height>\textheight\textheight\else\Gin@nat@height\fi}
\makeatother
\setkeys{Gin}{width=\maxwidth,height=\maxheight,keepaspectratio}
\makeatletter
\def\fps@figure{htbp}
\makeatother

\makeatletter
\@ifpackageloaded{caption}{}{\usepackage{caption}}
\AtBeginDocument{%
\ifdefined\contentsname
  \renewcommand*\contentsname{Table of contents}
\else
  \newcommand\contentsname{Table of contents}
\fi
\ifdefined\listfigurename
  \renewcommand*\listfigurename{List of Figures}
\else
  \newcommand\listfigurename{List of Figures}
\fi
\ifdefined\listtablename
  \renewcommand*\listtablename{List of Tables}
\else
  \newcommand\listtablename{List of Tables}
\fi
\ifdefined\figurename
  \renewcommand*\figurename{Figure}
\else
  \newcommand\figurename{Figure}
\fi
\ifdefined\tablename
  \renewcommand*\tablename{Table}
\else
  \newcommand\tablename{Table}
\fi
}
\@ifpackageloaded{float}{}{\usepackage{float}}
\floatstyle{ruled}
\@ifundefined{c@chapter}{\newfloat{codelisting}{h}{lop}}{\newfloat{codelisting}{h}{lop}[chapter]}
\floatname{codelisting}{Listing}

\makeatother
\makeatletter
\@ifpackageloaded{caption}{}{\usepackage{caption}}
\@ifpackageloaded{subcaption}{}{\usepackage{subcaption}}
\makeatother

\ifLuaTeX
  \usepackage{selnolig}  
\fi
\usepackage[]{natbib}
\usepackage{bookmark}

\IfFileExists{xurl.sty}{\usepackage{xurl}}{} 
\hypersetup{
  pdftitle={Title},
  pdfauthor={Author 1; Author 2},
  pdfkeywords={3 to 6 keywords, that do not appear in the title},
  colorlinks=true,
  linkcolor={blue},
  filecolor={Maroon},
  citecolor={Blue},
  urlcolor={Blue},
  pdfcreator={LaTeX via pandoc}}

\newcommand{\anon}{1}

\usepackage{amsthm}
\usepackage{enumerate}
\usepackage{enumitem}
\usepackage{booktabs}
\usepackage{algorithm}
\usepackage{algorithmic}
\usepackage{bm}
\newtheorem{theorem}{Theorem}
\newtheorem{proposition}{Proposition}

\DeclareMathOperator*{\argmax}{\arg\!\max}
\DeclareMathOperator*{\argmin}{\arg\!\min}

\newcommand{\A}{\mathcal{A}}

\newcommand{\C}{\mathbb{C}}

\allowdisplaybreaks

\begin{document}

\def\spacingset#1{\renewcommand{\baselinestretch}%
{#1}\small\normalsize} \spacingset{1}


\if1\anon
{
  \title{\bf Robust Estimation of Location in Matrix Manifolds Using the Projected Frobenius Median}
  \author{Houren Hong\thanks{The authors gratefully acknowledge that this research was supported by Australian Research Council grant DP220102232.\\
    Corresponding author's email: houren.hong@anu.edu.au}, 
    Kassel Liam Hingee\footnotemark[1], 
    Janice L. Scealy\footnotemark[1] \\
    and Andrew T.A. Wood\footnotemark[1]~
    \hspace{.2cm}\\
    Research School of Finance, Actuarial Studies and Statistics,\\
    Australian National University\\
    }
  \maketitle
} \fi

\if0\anon
{
  \bigskip
  \bigskip
  \bigskip
  \begin{center}
    {\LARGE\bf Robust Estimation of Location in Matrix Manifolds Using the Projected Frobenius Median}
\end{center}
  \medskip
} \fi

\bigskip
\begin{abstract}
We propose a robust method for location estimation in various matrix manifolds based on the projected Frobenius median, which is closely related to the spatial median. This method applies broadly to matrix manifolds, including Stiefel and Grassmann manifolds, Kendall shape spaces as well as to projective Stiefel manifolds, a type of quotient space of a Stiefel manifold. Our approach involves computation of the Frobenius median in an ambient Euclidean space followed by projection onto the relevant matrix manifold. Our estimation method is computationally attractive, has a unique solution provided the sample data are not colinear in the ambient Euclidean space, has desirable robustness features and has appropriate equivariance properties under natural groups of transformations. We establish asymptotic normality under mild conditions and derive the influence function for matrix manifolds of interest. Simulation studies on the rank-$1$ complex Grassmann manifold and the projective Stiefel manifold further show the applicability and robustness of our method. We also apply our method to a real-world earthquake moment tensor dataset. 
\end{abstract}

\noindent%
{\it Keywords:} extrinsic distance; Frobenius norm; influence function; spatial median; non-Euclidean spaces
\vfill

\newpage
\spacingset{1.8} 

\section{Introduction}\label{sec:intro}

\subsection{Background}

Robust estimators of location are important and valuable in statistics and have many uses, not only with traditional univariate and multivariate data, but also with data whose structure is more complicated, e.g. manifold data.  Our interest here is in matrix-valued data, specifically matrix data which lies in curved matrix manifolds. Current robust location estimators in this setting suffer from non-uniqueness, premature convergence (around local minima) and sensitivity to tuning parameters \citep{pennec2006intrinsic, fletcher2008robust, lin2024robust}.
Here we propose a new type of location estimator, the projected Frobenius median (PFM), that reliably produces robust location estimates in a wide variety of matrix manifolds.  PFMs have some excellent properties including: uniqueness; they are highly robust; when appropriately implemented, they have desirable equivariance properties;  and they are simple to calculate using well-established software for the spatial median.
Our numerical results indicate that PFMs will prove very useful in many contexts, e.g. for $k$-sample location problems in the presence of outliers, provided estimators of relevant covariances are robust.

Data on matrix manifolds arise naturally in a wide variety of applications. For example, surveys on the use of Stiefel and Grassmann manifolds in computer vision are provided in the work of \citet{turaga2008statistical} and \citet{lui2012advances}, while rank-$1$ complex Grassmann manifolds play an important role in statistical shape analysis \citep{kent1994complex,goodall1999projective,dryden2016statistical}.
Real symmetric non-negative definite matrices of fixed rank are closely connected to the estimation of covariance matrices, with applications ranging from diffusion tensor imaging \citep{dryden2009non,dryden2010power}, the analysis of networks data via graph Laplacians \citep{severn2022manifold,zhou2022network} to low-rank matrix completion \citep{candes2010matrix, chen2020noisy,wang2025robust}, among others.
Another important matrix space of interest is the projective Stiefel manifold, a quotient space of a Stiefel manifold in which the orthogonal columns are unsigned unit vectors. These spaces arise in a range of applications in statistics, including the statistics of directions and orientations, statistical shape analysis 
and computer vision \citep{arnold2013statistics, arnold2018statistics}. Despite their broad applicability, projective Stiefel manifolds have received relatively little attention in the statistical literature.

A natural and widely used intrinsic robust estimator is the Fr\'echet median, also known as the geometric median, considered by e.g.  \cite{pennec2006intrinsic} and \cite{fletcher2008robust}.  However, while conceptually appealing,  the use of intrinsic distance is typically far more computationally challenging than the approach proposed here, due in large part to the lack of uniqueness of the estimator.
Building on the idea of the Fr\'echet median,  \citet{lin2024robust} extend the median-of-means (MoM) strategy to manifold-valued data, motivated by the fact that the geometric median of a set of weakly concentrated estimators provides improved deviation bounds in a Hilbert space \citep{minsker2015geometric}. 
However, the computations of these approaches are generally iterative \citep{fletcher2008robust}, owing to minimising a sum of intrinsic distances, and on some matrix manifolds (e.g.~Stiefel manifolds) the intrinsic distance does not even have a closed-form expression.
The uniqueness and robustness of the MoM approach in the presence of outliers remain underexplored.  
Further, the performance of the MoM method largely hinges on the number of subsets and their partitioning.

\subsection{The projected Frobenius median}

In the context of directional statistics, \citet{ducharme1987spatial} proposed a robust estimator, $\hat{\mu}_{\textrm{DM}}$ say, of location on the sphere, given a sample of unit vectors $x_1, \ldots , x_n \in \mathbb{S}^{p-1} \subset \mathbb{R}^p$. This estimator, which is also discussed by \citet{chan1993median} and \citet{ko1993robust}, is defined as follows: first, calculate the spatial median, $\hat{\mu}_{\textrm{SM}}$ say, of the $x_i$ in the ambient space $\mathbb{R}^p$ (the spatial median is defined in (\ref{spatial_median})); and then normalize $\hat{\mu}_{\textrm{SM}}$ to obtain $\hat{\mu}_{\textrm{DM}}=\hat{\mu}_{\textrm{SM}}/\vert \vert \hat{\mu}_{\textrm{SM}} \vert \vert_E$, where $\vert \vert \cdot \vert \vert_E$ denotes the Euclidean norm. This estimator has some excellent properties: it is unique provided the $x_i$ are not all colinear; it is straightforward to calculate using publicly available software for the spatial median; and it is highly robust.

It turns out that essentially the same idea can be applied to a broad range of matrix manifolds: we first find the Frobenius median (cf. the spatial median, but using the Frobenius norm defined in (\ref{Frobenius_norm})) in a suitable ambient linear matrix space containing the matrix manifold of interest; and then we project the Frobenius median onto this matrix manifold to obtain the PFM.  
However, the practical details for matrix manifolds are rather more complicated than for the sphere.  One such complication, already mentioned, is the need to project from the relevant ambient linear matrix space to the matrix manifold of interest (on the sphere, projection is equivalent to normalization and the situation is far simpler on the sphere than in matrix manifolds).  Another challenge is to determine the influence function and central limit theorem for the PFM in a variety of matrix spaces.  These and other questions are systematically and thoroughly addressed in this paper.  We show that the resulting estimator is straightforward to use and has excellent theoretical and practical properties.

The use of an extrinsic distance (specifically,  the Frobenius norm), and performing most calculations in the ambient space, leads to a far simpler computational procedure than would any procedure that minimizes a sum of intrinsic distances. Moreover, as we shall see, the Frobenius median is essentially a (Euclidean) spatial median, but we use the name Frobenius median to emphasize the matrix structure.  A key point is that the excellent properties of the spatial median and its projected version are inherited by the Frobenius median and its projected version.

In this paper we focus on four matrix spaces: (i) the real Stiefel manifold; (ii) the real  Grassmann manifold; (iii) complex projective space, which for convenience is represented as a rank-$1$ complex Grassmann manifold and is important in statistics as the Kendall shape space for labelled landmark data for objects in two dimensions \citep{dryden2016statistical}; and (iv) projective Stiefel manifolds, defined below.  There are also many other types of matrix manifolds to which the PFM can conveniently be applied and relevant theory developed; such matrix manifolds are briefly listed in Section 2.5.

Why is robustness potentially an important issue for the sphere or for the matrix manifolds we consider in this paper, bearing in mind that these are all compact spaces?  The answer is that, even for compact spaces, robustness can be a serious issue if the data are at least moderately concentrated, and especially if they are highly concentrated.  A practical situation in matrix manifolds where robustness may be an issue is for a $k$-sample test of equality of locations for concentrated data when outliers are present.



\subsection{Outline of the paper}

 In Section 2 we specify the matrix manifolds that we focus on in this paper, and in Section 2.5 we list many other matrix manifolds to which the ideas and methods of this paper can be applied.  In Section 3 we give a precise definition of the projected Frobenius median and explicitly define the projections we need for the matrix manifolds considered in this paper.  We present our main theoretical results in Section 4; specifically, we give the form of the influence function and the central limit theory for Stiefel and Grassmann manifolds and complex projective space, while in Section 5 we present the results of simulation studies and also the analysis of a real earthquake moment tensor dataset. Moreover, we provide proofs of all results stated in the paper, plus the statements and proofs of some further results, in Section S1 of the Supplementary Material; some algorithms are presented in Section S2; and further numerical results are presented in S3.

\section{Matrix manifolds of interest}


\subsection{Stiefel manifolds}

Suppose that $X \in \mathbb{R}^{k \times r}$, i.e. $X$ is a real $k \times r$ matrix, where $\mathbb{R}$ is the set of real numbers.  Without loss of generality it is assumed that $1 \leq r \leq k$.  The real Stiefel manifold, denoted $\mathcal{V}_{k,r}$,
is defined by
\begin{equation}
\mathcal{V}_{k,r}=\left \{ X \in \mathbb{R}^{k \times r}: \, X^\top X=I_r    \right \},
\label{Stiefel_manifold}
\end{equation}
where $I_r$ is the $r \times r$ identity matrix.  Note that when $r=1$, $X$ is a $k \times 1$ unit vector and $\mathcal{V}_{k,1}=\mathcal{S}^{k-1}$, where $\mathcal{S}^{k-1}=\{x \in \mathbb{R}^k:\, x^\top x=1\}$ is the $p$-dimensional unit sphere.  Note also that, when $r=k$, the corresponding Stiefel manifold is the space of $k \times k$ orthogonal matrices.  Stiefel manifolds arise when there is an interest in the orientation of objects; see \citet{mardia2000directional}.  

\subsection{Grassmann manifolds}

The real Grassmann manifold, which intuitively is the space of subspaces of dimension $r$ in $\mathbb{R}^k$, is denoted $\mathcal{G}_{k,r}$, and may be defined by
\begin{equation}
\mathcal{G}_{k,r}=\left \{ Y \in \mathbb{R}^{k \times k}:\, Y^\top = Y = Y^2, \textrm{tr}(Y)=r\right \},
\label{Grassmann_manifold}
\end{equation}
where $\textrm{tr}(\cdot)$ denotes trace and $Y^\top$ is the transpose of $Y$. Note that, in words, $\mathcal{G}_{k,r}$ consists of the $k \times k$ projection matrices.  In the case $r=1$, $\mathcal{G}_{k,1}$ is the $k-1$ dimensional real projective space, denoted $\mathcal{RP}^{k-1}$, corresponding to unsigned directions. See \citet{mardia2000directional}.  Grassmann manifolds arise, for example, in computer vision.

\subsection{Complex projective space}

Complex projective space is classical in differential geometry.  In statistics, it is best known as being the shape space, defined by D. Kendall,  for objects in two dimensions described by labelled landmarks; see \citet{kendall1999shape} and \citet{dryden2016statistical}. Complex projective space, denoted $\mathcal{CP}^{k-1}$, maybe conveniently defined as follows via what is known as  the Veronese-Whitney embedding defined in e.g. \citet[p. 112]{dryden2016statistical}:
\begin{equation}
\mathcal{CP}^{k-1} = \left \{ Z \in \mathbb{C}^{ k\times k}: \, Z^\ast = Z = Z^2, \, \textrm{tr}(Z)=1 \right \},
\label{Complex_projective_space}
\end{equation}
where $\mathbb{C}$ denotes the set of complex real numbers, $Z^\ast$ is the conjugate transpose of a complex matrix $Z$, and $Z \in \mathcal{CP}^{k-1}$ is a Hermitian matrix (i.e. $Z^\ast = Z$) with a single non-zero eigenvalue equal to $+1$.  

\subsection{Projective Stiefel manifolds}

Define $\mathcal{E}_r=\{ \epsilon=(\epsilon_1, \ldots , \epsilon_r): \epsilon_j = \pm 1, \, j=1, \ldots , r\}$. Note that $\mathcal{E}_r$ has the structure of a group when we define the group operation $*$ for $\epsilon, \eta \in \mathcal{E}_r$ by
\[
\epsilon* \eta =(\epsilon_1, \ldots , \epsilon_r)*(\eta_1, \ldots , \eta_r)=(\epsilon_1 \eta_1, \ldots , \epsilon_r \eta_r) \in 
\mathcal{E}_r.
\]
For $X=[x_1, \ldots , x_r] \in \mathcal{V}_{k,r}$, define the group action of $\mathcal{E}_r$ on $\mathcal{V}_{k,r}$ by 
\begin{equation}
(\epsilon, X) \mapsto X_\epsilon=[\epsilon_1 x_1, \ldots , \epsilon_r x_r].
\label{Group_action_X}
\end{equation}
Note that $X_\epsilon \in \mathcal{V}_{k,r}$ for all $\epsilon \in \mathcal{E}_r$ since the columns of $X$ remain orthonormal when multiplied by $\pm 1$.
Formally, the projective Stiefel manifold, written $\mathcal{PV}_{k,r}$, is the quotient space
$\mathcal{PV}_{k,r}=\mathcal{V}_{k,r}/\mathcal{E}_r$.
Intuitively, this means that we identify all $2^r$ elements of $\mathcal{V}_{k,r}$  of the form $X=[\pm x_1, \ldots , \pm x_r]$ as being the same element of $\mathcal{PV}_{k,r}$. Note that $\mathcal{PV}_{k,1}=\mathcal{G}_{k,1}$.


%
There has been some consideration of related but different quotient Stiefel manifolds in the mathematics literature, see e.g. \citet{gitler1968projectivea,gitler1968projectiveb,basu2017topology}.
In the statistics literature, so far as we are aware, the first substantial consideration of projective Stiefel manifolds is given in \citet{arnold2013statistics}; see also \citet{kume2013saddlepoint} for related computational work.



\subsection{Other matrix manifolds of potential interest} \label{different_matrix_manifolds}

We mention, without going into details, a number of other manifolds of potential or actual practical interest for which similar methods to those proposed later in this paper may be developed.  These are:
(a) robust rank $r$ location estimators for real symmetric $k \times k$ matrices; (b) robust rank $r$ non-negative definite location estimators for real, symmetric non-negative definite $k \times k$ matrices; (c) cases (a) and (b) with additional eigenvalue constraints such as $\sum_{j=1}^r \lambda_j = 0$ and/or $\sum_{j=1}^r \lambda_j^t=1$ for $t \geq 1$, especially $t=1, 2$; (d) robust rank $r$ location estimators for real $k \times s$ matrices, where $1 \leq r \leq s$; (e) robust rank $r$ location matrices for complex Hermitian versions of the Grassmann manifold and also for cases (a)-(c) above; and
(f) complex versions of Stiefel manifolds, projective Stiefel manifolds and case (d) above.

 Robust location estimation in setting of type (b) and (c) arises in \citet{dryden2008multi}, where, respectively, $(b)$ corresponds to the notion of size-and-shape and (c) corresponds to the notion of shape developed in that paper.  Case (e) arises in the rank-$1$ case of complex Grassmann manifolds which corresponds to complex projective space. We are not aware of specific applications of type (a), (d), or (f) but we believe that (a) and (d) in particular are likely to be useful in some contexts.

 The key point is that the projected Frobenius median can be defined and computed in all of these spaces.  Details of the results for the Frobenius median in the spaces listed above are broadly similar to those for the spaces considered in detail in this paper.

\section{The Frobenius median on matrix manifolds}

\subsection{The main idea}

Let $\mathcal{M}$ denote one of the real matrix manifolds $\mathcal{V}_{k,r}$ or $\mathcal{G}_{k,r}$, or the complex matrix manifold $\mathcal{CP}^{k-1}$; see (\ref{Stiefel_manifold}), (\ref{Grassmann_manifold}) and (\ref{Complex_projective_space}) respectively. 
In all cases it is assumed that we observe a random sample $X_1, \ldots, X_n$ from $\mathcal{M}$ and we wish to determine a robust estimator of location in $\mathcal{M}$.

Recall that the Frobenius norm on the real matrix space $\mathbb{R}^{k \times t}$ is given by
\begin{equation}
\vert \vert X \vert \vert_F =\left \{ \textrm{tr}\left ( X^\top X \right ) \right \}^{1/2};
\label{Frobenius_norm}
\end{equation}
if we replace the real matrix space by the complex matrix space $\mathbb{C}^{k \times t}$, the Frobenius norm is similar to (\ref{Frobenius_norm}) but with the transpose in (\ref{Frobenius_norm}) replaced by conjugate transpose.
The population and sample Frobenius median in $\mathbb{R}^{k \times r}$ are defined, respectively, by
\begin{equation}
A_0=\argmin_{A \in \mathbb{R}^{k \times r}}=
\mathbb{E}_{X \sim F_0}\left [\vert \vert X-A \vert \vert_F \right], \hskip 0.3truein
\hat{A} = \argmin_{A \in \mathbb{R}^{k \times r}}\sum_{i=1}^n \vert \vert X_i-A\vert \vert_F.
\label{Sample_Frob_median}
\end{equation}

In what follows, we refer to the ambient space, denoted by $\mathcal{A}$,  of $\mathcal{M}$ as being $\mathbb{R}^{k \times r}$, $\mathbb{R}^{k \times k}$ and $\mathbb{C}^{k \times k}$,  when $\mathcal{M}$ is equal to $\mathcal{V}_{k,r}$, $\mathcal{G}_{k,r}$ and $\mathcal{CP}^{k-1}$, respectively.  
Our robust estimator of location in any of the matrix spaces $\mathcal{M}$ considered in Section 2 is based on the following two-step procedure.
\begin{description}
\item[(i)] Calculate the Frobenius median, $\hat{A} \in \mathcal{A}$, where $\mathcal{A}$ is the ambient space for $\mathcal{M}$.
\item[(ii)] Project $\hat{A}$ onto $\mathcal{M} \subset \mathcal{A}$, to obtain the projected Frobenius median , $\hat{M} \in \mathcal{M}$.
\end{description}
A similar two-step procedure has been used previously by \citet{ducharme1987spatial} to construct a robust estimator of location on the sphere.  However, so far as we are aware, this simple but convenient and practically useful approach has not previously been exploited systematically to obtain a robust location estimator in a class of matrix manifolds.
An attractive feature of the resulting robust estimator $\hat{M} \in \mathcal{M}$ is that this estimator is unique (except in very special circumstances) and straightforward to compute.

\subsection{Calculation of Frobenius median}

From a practical point of view, it is convenient to express the Frobenius norm in terms of the Euclidean norm. This is because there are efficient and reliable programs for calculating the Euclidean spatial median. If $y_1, \ldots , y_n$ is a sample of vectors in $\mathbb{R}^p$, then the sample spatial median is defined by
\begin{equation}
\hat{m} = \argmin_{m \in \mathbb{R}^p} \sum_{i=1}^n \vert \vert y_i - m \vert \vert_E ,
\label{spatial_median}
\end{equation}
where $\vert \vert y \vert \vert_E=(y^\top y)^{1/2}$ is the standard Euclidean norm.  

In the case $\mathcal{M}=\mathcal{V}_{k,r}$,  we use the standard vectorization operator $\textrm{vec}(\cdot)$; see e.g. \cite{mardia1979multivariate}.  Then it is easy to see that 
\begin{equation}
\vert \vert X \vert \vert_F=\vert \vert \textrm{vec}(X) \vert \vert_E.
\label{firsty}
\end{equation}

The situation is slightly different in the case $\mathcal{M}=\mathcal{G}_{k,r}$, because here we are dealing with symmetric matrices.  In this case, we use a vectorization operator, referred to below as $\textrm{vech}_{\sqrt{2}}(\cdot)$, that is different to $\textrm{vec}(\cdot)$ in two respects: first, it only vectorizes the lower triangle of the matrix, including the diagonal; and second, all off-diagonal elements are multiplied by $\sqrt{2}$. In the case of a symmetric matrix $A=(a_{ij})_{i,j=1}^3$, 
\begin{equation}
\textrm{vech}_{\sqrt{2}}(A)=\left (a_{11}, \sqrt{2} a_{21}, \sqrt{2}a_{31},
a_{22}, \sqrt{2}a_{32}, a_{33} \right )^\top.
\label{Examples}
\end{equation}
Then it is straightforward to check that, for any symmetric matrix $A$,
\begin{equation}
\vert \vert A \vert \vert_F =\vert \vert \textrm{vech}_{\sqrt{2}}(A) \vert \vert_E.
\label{secondy}
\end{equation}

Finally, in the case $\mathcal{CP}^{k-1}$, we use a vectorization operator appropriate for Hermitian matrices. Let $A=U + iV$, with $i = \sqrt{-1}$, denote a $k \times k$ Hermitian matrix.  Then $A^\ast=U^\top-iV^\top=U+iV=A$, where ${}^\ast$ denotes conjugate transpose. It is seen that the real part of $A$, $U=\mathcal{R}\textit{e}(A)$, is symmetric (i.e. $U^\top =U$), while the imaginary part of $A$, $V=\mathcal{I}\textit{m}(A)$, is skew-symmetric (i.e. $V^\top =-V$).  Let $\textrm{vecl}(V)$ denote the lower triangle of $V$ with the diagonal excluded; in the example (\ref{Examples}), $\textrm{vecl}(V)=(a_{21}, a_{31}, a_{32})^\top$.  Then an appropriate vectorization operator, $\textrm{vec}_{\mathcal{H}}(A)$, with $\mathcal{H}$ for Hermitian,  may be written
\begin{equation}
\textrm{vec}_{\mathcal{H}}(A)
=\left ( \textrm{vech}_{\sqrt{2}}(\mathcal{R}\textit{e}(A))^\top, \sqrt{2} \textrm{vecl}(\mathcal{I}\textit{m}(A))^\top   \right )^\top,
\label{vec_Hermitian}
\end{equation}
where $\textrm{vech}_{\sqrt{2}}(\cdot)$ and $\textrm{vecl}(\cdot)$ is defined in (\ref{Examples}) and above (\ref{vec_Hermitian}), respectively; and then
\begin{equation}
\vert \vert A \vert \vert_F = \sqrt{\textrm{tr}\left ( A^\ast A  \right )}
= \vert \vert \textrm{vec}_{\mathcal{H}}(A) \vert \vert_E.
\label{thirdy}
\end{equation}

Proofs of claims (\ref{firsty}), (\ref{secondy}) and (\ref{thirdy}) are given in the Supplementary Material.

\subsection{The projection step}

It turns out that for many matrix manifolds of interest, the projection step may be expressed in terms of quantities in the relevant spectral decomposition or SVD.  We summarize the projection results for the Stiefel and Grassmann manifolds and complex projective space in Proposition 1 below.  For a matrix manifold $\mathcal{M}$, let $\pi(A;\mathcal{M})$ denote the projection of $A \in \mathcal{A}$ onto $\mathcal{M}$, where $\mathcal{A}$ is a convenient ambient matrix space of $\mathcal{M}$.

\begin{proposition}\label{p:projection}
\noindent (i)  Consider $\mathcal{M}=\mathcal{V}_{k,r}$, the Stiefel manifold.  Suppose the sample Frobenius median $\hat{A}_s$ has the singular value decomposition (SVD) $\hat{A}_s=\sum_{j=1}^r \hat{\rho}_j \hat{s}_j \hat{t}_j^\top$,
where $\hat{\rho}_1 > \cdots > \hat{\rho}_r \geq 0$, the $k \times 1$ vectors  $\hat{s}_1, \ldots , \hat{s}_r$ are orthonormal and the $r \times 1$ vectors $\hat{t}_1, \ldots , \hat{t}_r$ are orthonormal.  Then 
$\hat{M}_s=\pi(\hat{A}_s; \mathcal{V}_{k,r})=\sum_{j=1}^r \hat{s}_j \hat{t}_j^\top$ is the unique projection of $\hat{A}_s$ onto $\mathcal{V}_{k,r}$.

\noindent (ii)  Consider $\mathcal{M}=\mathcal{G}_{k,r}$, the Grassmann manifold. Assume that the sample Frobenius median $\hat{A}_g \in \mathcal{A}$, has spectral decomposition
$\hat{A}_g=\sum_{j=1}^k \hat{\lambda}_j \hat{q}_j \hat{q}_j^\top$,
where $\hat{\lambda}_1 \geq \cdots \geq \hat{\lambda}_k$, with $\hat{\lambda}_r > \hat{\lambda}_{r+1}$, and $\hat{q}_1, \ldots , \hat{q}_k$ an orthonormal set of $k \times 1$ vectors. Then
$\hat{M}_g=\pi(\hat{A}_g; \mathcal{G}_{k,r})=\sum_{j=1}^r \hat{q}_j \hat{q}_j^\top$ is the unique projection of $\hat{A}_g$ onto $\mathcal{G}_{k,r}$.

\noindent (iii) Consider the complex projective space $\mathcal{M}=\mathcal{CP}^{k-1}$.  Suppose the sample Frobenius median, $\hat{A}_{c}$, has spectral decomposition
$\hat{A}_{c}=\sum_{j=1}^k \hat{\lambda}_j \hat{u}_j \hat{u}_j^\ast$,
where $\hat{\lambda}_1 > \hat{\lambda}_2 \geq \cdots \geq \hat{\lambda}_k$ and the $\hat{u}_j$ are complex unit vectors (i.e. $\hat{u}_j^\ast \hat{u}_j=1$ for each $j=1, \ldots , k$) such that $\hat{u}_i^\ast \hat{u}_j=0$ if $i \neq j$.  Then
$\hat{M}_c=\pi(\hat{A}_c; \mathcal{CP}^{k-1})=\hat{u}_1 \hat{u}_1^\ast$ is the unique projection of $\hat{A}_c$ onto $\mathcal{CP}^{k-1}$.
\end{proposition}

The notation $\hat{A}_s$, $\hat{A}_g$, $\hat{A}_c$, $\hat{M}_s$, $\hat{M}_g$ and $\hat{M}_c$ used in Proposition 1 will be used below.  

\subsection{Equivariance properties of the projected Frobenius median}  The projected Frobenius median for the matrix spaces $\mathcal{V}_{k,r}$, $\mathcal{G}_{k,r}$ and $\mathcal{CP}^{k-1}$  have  desirable and natural equivariance properties.  Note that these properties hold without any requirements being placed on the eigenvalues or singular values on the class of matrices of interest.  In each case it is assumed we observe a sample of matrices $X_1, \ldots , X_n$.

\begin{proposition}\label{p:equivariant}

\noindent (i) \textrm{The Stiefel manifold $\mathcal{V}_{k,r}$}.  For any $r \times r$ orthogonal matrix $R$ and any $k \times k$ orthogonal matrix $Q$, define the transformation $X_i \mapsto Q X_i R^\top$, $i=1, \ldots , n$.  
Then $\hat{M}_s=\pi(\hat{A}_s; \mathcal{V}_{k,r})$ transforms according to $\hat{M}_s \mapsto Q \hat{M}_s R^\top$.


\noindent (ii) \textrm{The Grassmann manifold $\mathcal{G}_{k,r}$}.  For any $k \times k$ orthogonal matrix $Q$, define  the transformation $X_i \mapsto Q X_i Q^\top $, $i=1, \ldots , n$. Then $\hat{ M}_g=\pi(\hat{A}_g; \mathcal{G}_{k,r})$ transforms 
according to $\hat{M}_g \mapsto Q \hat{M}_g Q^\top$.

\noindent (iii) \textrm{Complex Projective space $\mathcal{CP}^{k-1}$}.  For any $k \times k$ unitary matrix $U$, for the $X_i \in \mathcal{CP}^{k-1}$ given by  (\ref{Complex_projective_space}), define $X_i \mapsto U X_i U^\ast $, $i=1, \ldots , n$.  Then $\hat{M}_c=\pi(\hat{A}_c; \mathcal{CP}^{k-1})$ transforms 
according to $\hat{M}_c \mapsto U \hat{M}_c U^\ast$.

\end{proposition}

Note that the equivariances in Proposition 2(ii) and 2(iii) would not hold if we did not multiply off-diagonal elements by $\sqrt{2}$ as in (\ref{Examples}) and (\ref{vec_Hermitian}).

\subsection{Projective Stiefel manifolds}

Estimation and projection on a projective Stiefel manifold require care because we must account for both the sign ambiguity and the Stiefel manifold constraints.  We first define two nested ambient spaces: the first, $\mathcal{B}_{0}$, is defined by
\begin{equation}
\mathcal{B}_{0}=\left \{B=(B_1, \ldots , B_r): B_j^\top =B_j \hskip 0.1truein \textrm{with} \hskip 0.1truein  B_j \hskip 0.1truein k \times k, \hskip 0.1truein j=1, \ldots , r \right \};
\label{ambient_space_4_0}
\end{equation}
and the second ambient space is
\begin{equation}
\mathcal{B}_{1}=\left \{ B=(B_1, \ldots , B_r) \in \mathcal{B}_{0}: \hskip 0.1truein \textrm{rank}(B_j)=1, \hskip 0.1truein j=1, \ldots , r      \right \}
\label{ambient_space_4_1}
\end{equation}
If $X =[\pm x_1 , \ldots , \pm x_r] \in \mathcal{PV}_{k,r}$ has a population distribution $F_0$, then the population Frobenius median in the ambient space $\mathcal{B}_{0}$ is defined by
\begin{equation}
B_0=(B_{01}, \ldots , B_{0r})=\argmin_{B =(B_1, \ldots , B_r)\in \mathcal{B}_{0}}
\int \left \{\sum_{j=1}^r  \vert  \vert B_j - x_j x_j^\top  \vert  \vert_F^2 \right \}^{1/2}dF_0(X).
\label{Frob_med_in_A_4_0}
\end{equation}
The sample analogue of (\ref{Frob_med_in_A_4_0}), for a sample $X_1, \ldots , X_n \in \mathcal{PV}_{k,r}$, is written 
$\hat{B}=(\hat{B}_1, \ldots , \hat{B}_r)$.

We now specify two conditions on the population distribution $F_0$ through conditions on $B_0=(B_{01}, \ldots , B_{0r})$.  For each $j=1, \ldots , r$, assume that $B_{0j}$ has spectral decomposition $B_{0j}=\sum_{a=1}^k \lambda_{0ja} q_{0ja} q_{0ja}^\top$, where $\lambda_{0j1} \geq \cdots \geq \lambda_{0jk}$ and $q_{0j1}, \ldots , q_{0jk}$ are orthonormal vectors.  
\begin{equation}
\textrm{For} \hskip 0.1truein j=1, \ldots , r, \hskip 0.1truein \lambda_{0j1} > \lambda_{0j2}; \, \, \textrm{and}  \hskip 0.1truein q_{011}, \ldots , q_{0r1} \hskip 0.1truein \textrm{are linearly independent}.
\label{those_conditions}
\end{equation}

The following proposition defines a projection of $B \in \mathcal{B}_{0}$, defined in (\ref{ambient_space_4_0}), to $\mathcal{B}_{1}$, defined in (\ref{ambient_space_4_1}); this map is written $\pi(B; \mathcal{B}_{1})$.
\begin{proposition} \label{proj_stiefel_proj_1}
Suppose that $B=(B_1,\ldots, B_r) \in \mathcal{B}_{0}$, defined in (\ref{ambient_space_4_0}), and satisfies the conditions in (\ref{those_conditions}).  If $q_{j1}$ is the leading eigenvector for each $B_j$, then the projection $\pi(B; \mathcal{B}_{1})$ from $B \in \mathcal{B}_{0}$ to $\mathcal{B}_{1}$ is given by
\begin{equation}
\pi(B;\mathcal{B}_{1}) = \left(q_{11}q_{11}^\top,\ldots, q_{r1}q_{r1}^\top\right)
\in \mathcal{B}_{1}.
\label{proj4}
\end{equation}
\end{proposition}

Let $\epsilon=(\epsilon_1, \ldots , \epsilon_r)$ where $\epsilon_{j} = \pm 1$ and  write $\epsilon=1[r]$ when $\epsilon_1=\cdots=\epsilon_r=1$.
We now project $Q_{\epsilon}=[\epsilon_1 q_{11}, \ldots , \epsilon_r q_{r1}]$ to $\mathcal{V}_{k,r}$ and use this to construct an element of $\mathcal{PV}_{k,r}$. 
\begin{proposition}\label{proj_stiefel}
Suppose that $Q_{1[r]}=[q_{11}, \ldots , q_{r1}]$ has the SVD, 
$Q_{1[r]}= \sum_{j=1}^r \rho_j s_j t_j^\top$. 
Then the projection of $Q_\epsilon$ onto $\mathcal{V}_{k,r}$ is given by
\begin{equation}
 \pi(Q_\epsilon; \mathcal{V}_{k,r})=\sum_{j=1}^r  s_j t_j(\epsilon)^\top,
 \label{Q_epsilon_rep}
\end{equation}
where $t_j(\epsilon)=(\epsilon_1 t_{j1}, \ldots , \epsilon_r t_{jr})^\top$ and $t_j=(t_{j1}, \ldots , t_{jr})^\top$, $j=1, \ldots , r$.  
\end{proposition}
\noindent Therefore, to recover the entire coset $\{\pi(Q_\epsilon ; \mathcal{V}_{k,r}); \, \epsilon \in \mathcal{E}_r\}$, it is sufficient to identify a single representative such as $\pi(Q_{1[r]}; \mathcal{V}_{k,r})$, which determines the SVD unit vectors $s_j$ and $t_j$.  The cosets can then be generated through (\ref{Q_epsilon_rep}) by considering $\epsilon \in \mathcal{E}_r$, and the projected Frobenius median is thus identified.
The corresponding equivariance result is given.

\begin{proposition}
Let $X_1, \ldots , X_n \in \mathcal{PV}_{k,r} \subset \mathcal{V}_{k,r}$ denote a sample and let $Q$ denote a $k \times k$ orthogonal matrix and suppose $R=\textrm{diag}(\pm 1, \ldots , \pm 1)$ is $r \times r$.  Let $\hat{M}_{ps}$ denote any representative of the coset of size $2^r$ of sample projected Frobenius medians.  If we then apply the map $X_i \mapsto UX_i R$, $i=1, \ldots , n$, then the selected projected Frobenius median $\hat{M}_{ps}$ transforms according to $\hat{M}_{ps}  
\mapsto Q \hat{M}_{ps}$.
\end{proposition}

Note that (i) the projected Frobenius median does not depend on the sign matrix $R$; and (ii) this equivariance result is weaker than those in Proposition 2, due to the fact that two distinct projections are involved in calculating $\hat{M}_{ps}$.

\section{Influence functions and asymptotic normality}

We recall the definition of the influence function \citep[see e.g.][]{hampel1986robust}.  Let $F_0$ denote the population distribution on a sample space $\mathcal{X}$.  Consider a mixture distribution 
\begin{equation}
F_{z,\epsilon}(x)=(1-\epsilon) F_0(x) + \epsilon \delta_z(x), \hskip 0.2truein z \in \mathcal{X},
\label{basic_mixture}
\end{equation}
where $\delta_z(x)$ is the probability distribution that assigns probability 1 to $x=z$.  Let $T(F)$ denote a functional on $\mathcal{F}$, a large class of probability distributions on $\mathcal{X}$. The influence function for the estimator $T$ is the functional (Gateaux) derivative 
\begin{equation}
IF_{T;F_0}(z)=\lim_{\epsilon \downarrow 0} \left (\frac{T(F_{z,\epsilon})-T(F_0)}{\epsilon} \right ),
\label{Influence_function}
\end{equation}
where $F_{z,\epsilon}$ is defined in (\ref{basic_mixture}) and $F_{z,0}=F_0$; see e.g. \citet{hampel1986robust}.  As explained in Section 3.3, $\pi(\cdot;\cdot)$ denotes a general projection map in this section.

It is clear from the above that, for the three spaces considered, i.e. Stiefel and Grassmann manifolds and complex projective space, both the influence function and the central limit theorem for the projected Frobenius median have quite a lot of common structure. This common structure is also shared by many of the matrix spaces mentioned in Section 2.5.  However, some of the details in each case are different, which is why we have presented the three cases above separately. The results given in the Supplementary Material for the projective Stiefel manifold are also broadly similar but the formulas are more complicated due to there being two projections involved.   

In what follows, each of Theorems 1-3 is presented in the same format: in part (i), we derive the influence function of the Frobenius median for Stiefel and Grassmann manifolds and for complex projective space, respectively; and in part (ii), the corresponding central limit theorem is established, formulated in the relevant tangent space in each case.

\subsection{Stiefel manifolds}
Suppose $\mathcal{M}=\mathcal{V}_{k,r}$ is a Stiefel manifold with $r<k$.  Let $A_{0s}=\sum_{j=1}^r \rho_{0j} s_{0j} t_{0j}^\top$, with $\rho_{01} \geq \cdots \geq \rho_{0r}>0$, denote the singular value decomposition of $A_{0s}$.
The projected Frobenius median $M_{0s}=\pi(A_{0s};\mathcal{V}_{k,r})$ is given by $M_{0s}=\sum_{j=1}^r s_{0j} t_{0j}^\top$.  Define
\begin{equation}
H_s=\mathbb{E}_{X \sim F_0}
\left [ \frac{1}{\vert \vert \textrm{vec}(X-A_0 )\vert \vert_E} \left (I_{(kr) \times (kr)}-\frac{\textrm{vec}(X-A_{0s})\, \textrm{vec}(X-A_{0s})^\top}{\vert \vert X-A_{0s} \vert \vert_F^2}  \right ) \right ]
\label{H_1_tex}
\end{equation}
and
\begin{equation}
J_s=\mathbb{E}_{X \sim F_0}\left [  \frac{\textrm{vec}(X-A_{0s}) \textrm{vec}(X-A_{0s})^\top }{\vert \vert X-A_{0s} \vert \vert_F^2} \right ].
\label{J_1_tex}
\end{equation}
Let $s_{01}, \ldots , s_{0k}$ denote a set of orthonormal $k$-vectors, where $s_{01}, \ldots , s_{0r}$ are singular vectors of $A_{0s}$, as above.  If $\otimes$ denotes the Kroneckor product, define $\Xi_s=[\Xi_{1,s} ,\Xi_{2,s}]$, where
\begin{equation}
\Xi_{1,s}=\left [\xi_{1,s;ab}= \frac{1}{\sqrt{2}}(t_{0j}\otimes s_{0a} - t_{0a} \otimes s_{0b}  ):  1 \leq a < b \leq r \right ]
\label{xi_1_s}
\end{equation}
and
\begin{equation}
\Xi_{2,s}= [\xi_{2,s;ab}= t_{0a} \otimes s_{0b}:  1 \leq a \leq r<b\leq k].
\label{xi_2_s}
\end{equation}
 Note that the columns of $\Xi_s$ provide an orthonormal basis for the tangent space of $\textrm{vec}(\mathcal{V}_{k,r})$ at $\textrm{vec}(M_{0s})$, where $M_{0s} \in \mathcal{V}_{k,r}$.  Let $\hat{A}_s$ denote the ambient sample Frobenius median.

\begin{theorem}
\noindent (i) Using notation in (\ref{basic_mixture}) and (\ref{Influence_function}), $\hat{M}_s=\pi(\hat{A}_s; \mathcal{V}_{k,r})$ has influence function
\begin{align}
IF_{\hat{M}_s; F_0}(Z)&=\lim_{\epsilon \downarrow 0} \frac{M_s(Z, \epsilon) - M_{0s}}{\epsilon} \nonumber \\
&=\sum_{1 \leq a < b \leq r} \frac{1}{\rho_{0a}+ \rho_{0b}}
\left \{s_{0a}^\top IF_{\hat{A}_s;F_0}(Z) t_{0b} - s_{0b}^\top IF_{\hat{A}_s; F_0}(Z) t_{0a} \right \}\left \{  s_{0a} t_{0b}^\top -s_{0b}  t_{0a}^\top \right \} \nonumber \\
& \hskip 0.5truein + \sum_{a=1}^r \sum_{j=r+1}^k \frac{1}{\rho_{0j}} s_{0j}^\top IF_{\hat{A}_s;F_0}(Z)t_{0a} s_{0j} t_{0a}^\top ,
\label{IF_hat_M_s}
\end{align}
where $Z \in \mathcal{V}_{k,r}$, $I_{k \times k}$ is the $k \times k$ identity matrix  and
\begin{equation}
IF_{\hat{A}_s;F_0}(Z)=\lim_{\epsilon \downarrow 0} \frac{\left (\hat{A}_{Z, \epsilon}-A_{0s}\right )}{\epsilon}=
\textrm{vec}_{k \times r}^{-1}\left \{H_s^{-1} \frac{\textrm{vec}(Z-A_{0s})}{\vert \vert Z-A_{0s} \vert \vert_F } \right \},
\label{projected_Frobenius_A_s}
\end{equation}
where $H_s$  is defined in (\ref{H_1_tex}) and $\textrm{vec}_{k \times r}^{-1}$ is the inverse of the vec operator which maps a $(kr) \times 1$ vector to a $k \times r$ matrix.

\noindent (ii) Define $y_s =\Xi_s^\top \textrm{vec}(\hat{M}_s-M_0)$.  Then, as $n \rightarrow \infty$, $n^{1/2}y_s$ converges in distribution to 
$\mathcal{N}_{p}(0_p,C_s)$ where $p=kr-r(r+1)/2$ and $C_s$ has typical element
\begin{equation}
\textrm{Cov}(y_{\alpha, s; ab}, y_{\beta, s; cd})
=\frac{2^{4-\alpha -\beta}}{(\rho_{0a}+\rho_{0b})(\rho_{0c}+\rho_{0d})}\xi_{\alpha,s;ab}^\top V_s \xi_{\beta, s; cd}
\label{Stiefel_covariance}
\end{equation}
where $V_s=H_s^{-1} J_s H_s^{-1}$ with $H_s$ and $J_s$ defined in (\ref{H_1_tex}) and (\ref{J_1_tex}) respectively; for  $\alpha, \beta = 1,2$,  $y_{\alpha ,s; ab}=\xi_{\alpha , s;ab}^\top \textrm{vec}(\hat{M}_s - M_0)$, $\xi_{\alpha , s;ab}$ is defined in (\ref{xi_1_s}) ($\alpha=1$) or (\ref{xi_2_s}) ($\alpha=2$); and the $\rho_{0a}$ are the singular values of $A_{0s}$ with $\rho_{0a}=0$ for $r+1 \leq a \leq k$.  Moreover, if $V_s$ has full rank $kr$, then $C_s$ has full rank $p$.
\end{theorem}
When $k=r$, in Theorem 1, part (i), the only difference is that the second term on the RHS of (\ref{IF_hat_M_s}) is the relevant matrix of zeros; and in Theorem 1, part (ii), the only difference is that the matrix in (\ref{xi_2_s}) is the corresponding matrix of zeros, so that $\alpha=\beta=1$ in (\ref{Stiefel_covariance}).

Note that $p=kr-r(r+1)/2$ is actually the dimension of the Stiefel manifold $\mathcal{V}_{k,r}$.  Moreover, we may interpret $y_s$ as a vector of coordinates of a point in the tangent space at $\sum_{j=1}^r s_{0j} t_{0j}^\top \in \mathcal{V}_{k,r}$, where the coordinates are defined with respect to the orthonormal basis of the tangent space given by the columns of $\Xi_s$.

\subsection{Grassmann manifolds}
We now consider the influence function and asymptotic normality of the Frobenius median in a Grassmann manifold $\mathcal{G}_{k,r}$. Suppose that $A_{0g}=\sum_{j=1}^k \lambda_{0j} q_{0j} q_{0j}^\top$ denotes the population Frobenius median of the ambient population, where it is assumed that $\lambda_{01} \geq \cdots \geq \lambda_{0r}>\lambda_{0,r+1} \geq \cdots \geq \lambda_{0k}$.  
Let $\hat{A}_g$ denote the ambient sample Frobenius median and define 
\begin{equation}
H_g=\mathbb{E}_{X \sim F_0} \left [\frac{1}{\vert \vert X - A_{0g}\vert \vert_F} \left (I_{k(k+1)/2} - \frac{\textrm{vech}_{\sqrt{2}}(X-A_{0g})\, \textrm{vech}_{\sqrt{2}}(X-A_{0g})^\top}{\vert \vert X-A_{0g} \vert \vert_F^2} \right )  \right ],
\label{H_2}
\end{equation}
\begin{equation}
J_g=\mathbb{E}_{X \sim F_0} \left [  \frac{\textrm{vech}_{\sqrt{2}} (X-A_{0g}) \textrm{vech}_{\sqrt{2}}(X-A_{0g})^\top}{\vert \vert X-A_{0g} \vert \vert_F^2}   
\right ]
\label{J_2}
\end{equation}
 and the $\{k(k+1)/2\} \times \{k(k+1)/2\}$ matrix $G_k=\textrm{diag}(g_{t,k}: t=1, \ldots , k(k+1)/2)$, with
\begin{equation}
g_{t,k}=\begin{cases}
1 & \textrm{if} \hskip 0.1truein  t=(j-1)k + 1 - (k-1)(k-2)/2 \hskip 0.1truein \textrm{for some} \hskip 0.1truein
j=1, \ldots , k\\
\sqrt{2} &\textrm{otherwise}.
\end{cases}
\label{G_k}
\end{equation}

\begin{theorem}

\noindent (i) Consider $\hat{M}_g=\pi(\hat{A}_g; \mathcal{G}_{k,r})$ and suppose that the population distribution on $\mathcal{G}_{k,r}$ is $F_0$ and is such that $\lambda_{0r}>\lambda_{0,r+1}$.  Then the influence function for $\hat{M}_g$ is given by
\begin{align*}
IF_{\hat{M}_g;F_0}(Z)&=\lim_{\epsilon \downarrow 0} \frac{(M_g(Z, \epsilon) - M_{0g})}{\epsilon}\\
&=\sum_{1 \leq a \leq r} \sum_{r<b \leq k}
\frac{1}{\lambda_{0a} - \lambda_{0b}} \left (q_{0a}^\top IF_{\hat{A}_g;F_0}(Z) q_{0b} \right ) \left ( q_{0a} q_{0b}^\top + q_{0b} q_{0a}^\top \right ),
\end{align*}
where $Z \in \mathcal{G}_{k,r}$ and, writing $\textrm{vech}_{\sqrt{2}}(\cdot)^{-1}$ for the inverse operator of $\textrm{vech}_{\sqrt{2}}(\cdot)$ in (\ref{Examples}),
\[
IF_{\hat{A}_g; F_0}(Z)=\textrm{vech}_{\sqrt{2}}^{-1}\left (  H_g^{-1} \frac{\textrm{vech}_{\sqrt{2}}(Z-A_{0g})}{\vert \vert Z-A_{0g}\vert \vert_F}\right ).
\]

\noindent (ii) Define $y_g =\Xi_g^\top \textrm{vec}(\hat{M}_g-M_0)$ where the $k^2 \times \{r(k-r)\}$  matrix $\Xi_g$ is given by
\begin{equation}
\Xi_g=[\xi_{g;ab}= (q_{0a}\otimes q_{0b} + q_{0b} \otimes q_{0a})/\sqrt{2}: \, 1 \leq a \leq r < b \leq k].
\label{def_Xi_g}
\end{equation}
Then, as $n \rightarrow \infty$,  $n^{1/2} y_g$ converges in distribution to a $\mathcal{N}_\tau (0_\tau, C_g)$ distribution, where $\tau=r(k-r)$ and $C_g$ has full rank if $V_g$ has full rank.  A typical element of $C_g$ is given by
\begin{equation}
\textrm{Cov}(y_{g;ab}, y_{g;cd})=2(\lambda_{0a}-\lambda_{0b})^{-1} (\lambda_{0c} - \lambda_{0d})^{-1} \xi_{g;ab}^\top D_k G_k^{-1} V_g G_k^{-1} D_k^\top  \xi_{g;cd},
\label{equation}
\end{equation}
where $D_k$ is the $k^2 \times k(k+1)/2$ duplication matrix \citep[][p. 55]{magnus1988linear}, $G_k$ is defined in (\ref{G_k}), $V_g=H_g^{-1} J_g H_g^{-1}$, $y_{g;ab}=\xi_{g;ab}^\top \textrm{vec}(\hat{M}_g-M_0)$,
and the $\xi_{g;ab}$ is defined in (\ref{def_Xi_g}).
\end{theorem}
Note that $\tau=r(k-r)$ is the dimension of the Grassmann manifold $\mathcal{G}_{k,r}$.  Moreover, we may interpret $y_g$ as a vector of coordinates of a point in the tangent space at the point $\sum_{j=1}^r q_{0j} q_{0j}^\top \in \mathcal{G}_{k,r}$, where the coordinates are defined with respect to the orthonormal basis of the tangent space given by the columns of $\Xi_g$.

\subsection{Complex projective space}

The case where $\mathcal{M}=\mathcal{CP}^{k-1}$, complex projective space, is now considered.  This corresponds to the complex version of the Grassmann manifold with $r=1$.  Let $A_{0c}=\sum_{j=1}^k \lambda_{0j} u_{0j} u_{0j}^\ast $ where $\lambda_{01}> \lambda_{02} \geq \cdots \geq \lambda_{0k}$. 
Recalling the definition of $\textrm{vec}_{\mathcal{H}}(\cdot)$ in (\ref{vec_Hermitian}), define
\begin{equation}
H_c=\mathbb{E}_{Z \sim F_0} \left [\frac{1}{\vert \vert Z- A_{0c}\vert \vert_F} \left (I_{k^2 \times k^2} - \frac{\textrm{vec}_{\mathcal{H}}(Z-A_{0c})\, \textrm{vec}_{\mathcal{H}}(Z-A_{0c})^\ast}{\vert \vert Z-A_0 \vert \vert_F^2} \right )  \right ],
\label{H_3}
\end{equation}
\begin{equation}
J_c=\mathbb{E}_{Z \sim F_0} \left   [  \frac{\textrm{vec}_{\mathcal{H}}(Z-A_{0c}) \, \textrm{vec}_{\mathcal{H}}(Z-A_{0c})^\ast}{\vert \vert Z-A_{0c} \vert \vert_F^2} \right]
\label{J_3}
\end{equation}
and let $\hat{A}_c$ denote the ambient sample Frobenius median.

\begin{theorem}
\noindent (i) Consider $\hat{M}_c=\pi(\hat{A}_c; \mathcal{CP}^{k-1})$, let $F_0$ denote the population distribution on $\mathcal{CP}^{k-1}$ and suppose $F_0$ is such that $\lambda_{01}>\lambda_{02}$.  The influence function for $\hat{M}_c$ is given by
\[
IF_{\hat{M}_c;F_0}(Z)=\lim_{\epsilon \downarrow 0} \frac{(M_c(Z, \epsilon) - M_{0c})}{\epsilon}= \sum_{1<b \leq k}
\frac{1}{\lambda_{01} - \lambda_{0b}} \left (u_{01}^\ast IF_{\hat{A}_c;F_0}(Z) u_{0b} \right ) \left ( u_{01} u_{0b}^\ast + u_{0b} u_{01}^\ast \right ),
\]
where $Z \in \mathcal{CP}^{k-1}$ and, writing $\textrm{vec}_{\mathcal{H}}(\cdot)^{-1}$ for the inverse operator of $\textrm{vec}_{\mathcal{H}}(\cdot)$, which operates on real vectors of dimension $k^2$,
\[
IF_{\hat{A}_c;F_0}(Z)=\lim_{\epsilon \downarrow 0} \frac{(A_c(Z, \epsilon) - A_{0c})}{\epsilon}=\textrm{vec}_{\mathcal{H}}^{-1}\left (  H_c^{-1} \frac{\textrm{vec}_{\mathcal{H}}(Z-A_0)}{\vert \vert Z-A_0\vert \vert_F}\right ).
\]

\noindent (ii) Define $y_c =\Xi_c^\ast \textrm{vec}(\hat{M}_c-M_0)$ where the $k^2 \times (k-1)$ matrix $\Xi_c$ is given by
\begin{equation}
\Xi_c=[ \xi_{c;b}=(\bar{u}_{01}\otimes u_{0b}+ \bar{u}_{0b} \otimes u_{01})/\sqrt{2}: \, 1  < b \leq k], 
\label{Xi_c}
\end{equation}
where $\bar{u}_{0b}$ is the complex conjugate of $u_{0b}$.
Then, as $n \rightarrow \infty$,  $n^{1/2} y_c$ converges in distribution to a $\tau$-dimensional complex normal distribution $\mathcal{CN}_\tau (0_\tau, C_c)$, where $\tau=k-1$ and, provided $V_c$ has full rank, $C_c$ is a full rank Hermitian covariance matrix with typical element 
\[
\textrm{Cov}(y_{c;b}, \bar{y}_{c;d})=2(\lambda_{01} - \lambda_{0b})^{-1} (\lambda_{01} - \lambda_{0d})^{-1} \xi_{c;b}^\top \textrm{diag}\{D_kG_k^{-1}, \tilde{D}_k /\sqrt{2}\}V_c 
\textrm{diag}\{G_k^{-1} D_k^\top, \tilde{D}_k^\top \}\xi_{c;d}, 
\]
where $D_k$ is as in Theorem 2, $\tilde{D}_k$ is the duplication matrix for skew-symmetric matrices \citep[][p. 94]{magnus1988linear}, $V_c=H_c^{-1} J_c H_c^{-1}$, $y_{c;b}=\xi_{c;b}^\top \textrm{vec}(\hat{M}-M_0)$ and $\xi_{c;b}$ is defined in (\ref{Xi_c}).
\end{theorem}
Note that $\mathcal{C}^{k-1}$ has dimension $k-1$.  Moreover, $y_c$ is a vector of coordinates of a point in the tangent space at $u_{01} u_{01}^\ast \in \mathcal{CP}^{k-1}$, where the coordinates are defined with respect to the orthonormal basis of the tangent space given by the columns of $\Xi_c$.


\section{Numerical results}\label{sec:numerical}
\subsection{Simulation on planar shape space} 
\begin{figure}[ht]
\centering
\includegraphics[width=0.45\textwidth]{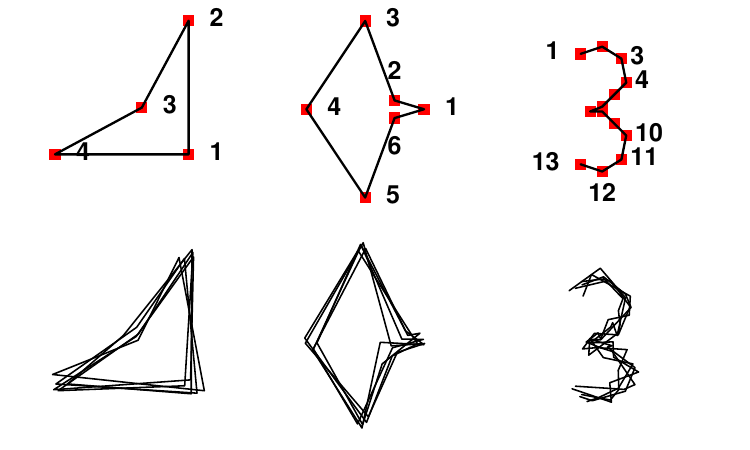}
\hfill
\includegraphics[width=0.5\textwidth]{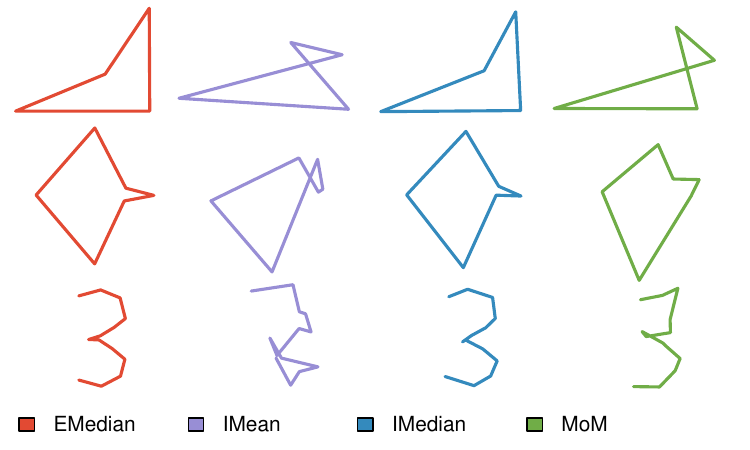}
\caption{Left: Illustrations of the original configurations (top row) and five samples from the complex Bingham distribution (bottom row). Right: Estimated configurations in the presence of $90$ outliers, obtained by EMedian (first column), IMean (second column), IMedian (third column) and MoM (last column).}
\label{fig:simu1_shape}
\end{figure}

Planar shape analysis has been widely studied in computer vision and medical imaging, focusing on the configuration of objects that is invariant to translation, scaling, and rotation. A planar configuration is commonly represented by $k\ge 3$ landmarks on its contour. These landmarks are encoded as a complex vector in $\C^k$, where each component corresponds to the Cartesian coordinates of a landmark.
In this section, we consider three configurations $c^{(j)}\in \C^k$ for $j=1,2,3$ as follows:
\begin{align*}
c^{(1)} & = (0.29-0.29i, 0.29+0.57i, -0.01+0.01i, -0.57-0.29i)^\top; \\
c^{(2)} & = (0.32+0.00i, 0.13+0.06i, -0.06+0.57i, -0.44+0.00i, \\
                &\qquad -0.06-0.57i, 0.13-0.06i)^\top ;\\
c^{(3)} & = (-0.17+0.36i, -0.03+0.41i, 0.10+0.33i, 0.13+0.17i,\\
                &\qquad  0.05+0.09i, -0.03+0.02i, -0.11-0.01i, -0.03-0.01i, \\
                & \qquad 0.05-0.09i, 0.13-0.17i, 0.10-0.33i, -0.03-0.40i, -0.17-0.36i)^\top.
\end{align*}
with their contours sketched in the left panel of Figure~\ref{fig:simu1_shape}. 

To make the original configurations invariant to translation and scaling, they are further transformed into \emph{pre-shapes}, denoted as $z^{(j)}$ for $j=1,2,3$; see also \citet[p.64]{dryden2016statistical}. This is achieved by $z^{(j)}=H c^{(j)}/\|H c^{(j)}\|\in \mathcal{CS}^{k-2}$ for $j=1,2,3$, where $H$ is a $(k-1)\times k$ Hermitian submatrix satisfying $H H^\top = I_{k-1}$ \citep[p.49]{dryden2016statistical}, and $\mathcal{CS}^{k-2}$ denotes the complex unit sphere. 
Given a pre-shape $z^{(j)}$ as a representative shape, the corresponding \emph{shape} is uniquely defined by the set $[z^{(j)}]=\{z^{(j)}e^{i\theta}:\theta\in[0, 2\pi]\}\in \mathcal{CP}^{k-2}$, where $i$ denotes the imaginary unit, $\theta$ is an anti-clockwise rotation angle about the origin and $\mathcal{CP}^{k-2}$ denotes the complex projective space \citep[p.82]{dryden2016statistical}. 

In what follows, we denote the representative shape for each configuration as $z_0\in \mathcal{CS}^{k-2}$. Moreover, as discussed in Section 2.3, it is more convenient to denote shapes via the Veronese-Whitney embedding. We therefore denote the population shape for each configuration as $Z_0 = z_0z_0^{\ast}\in\mathcal{CP}^{k-2}$, where $\ast$ is the conjugate transpose. 
The primary target of shape analysis is to recover $Z_0$ from samples that may be subject to contamination.

We generate $n=200$ independent and identically distributed data $\{x_j\}_{j=1}^{n}$ from a complex Bingham distribution for each configuration and embed these observations as matrices, i.e., $X_j = x_j x_j^{\ast}$ for $j=1,\ldots, n$. This probability density function of $x_j$ is proportional to $\exp(x^* \Lambda x)$, which is invariant under planar rotations $e^{i\theta}$; see also \citet[pp.326-334]{mardia2000directional}. Here, $\Lambda$ is a $(k-1)\times (k-1)$ Hermitian matrix with its eigenvalues controlling the concentration of the distribution. 
In this study, we set $z_0$ as the leading eigenvector of $\Lambda$, with the remaining eigenvectors obtained by Gram-Schmidt orthonormalization. The eigenvalues of $\Lambda$ for each case are $(150,0,0)^\top$, $(150,0,\ldots,0)^\top$ and $(200,0,\ldots,0)^\top$, respectively, so that the samples are highly concentrated around the population shape $Z_0=z_0 z_0^{\ast}$. Figure~\ref{fig:simu1_shape} (left bottom) displays five sample configurations drawn from the complex Bingham distribution for each case. 

To verify robustness, we randomly replace $n_1 = 20, 40, 90$ observations (10\%, 20\% and 45\% contamination) with outliers. These outliers are constructed as complex unit vectors which are orthogonal to $z_0$, and they correspond to directions that maximizes the distance from $z_0$ in $\mathcal{CP}^{k-2}$. Each outlier is generated as follows: first, a complex vector is drawn from the standard complex normal distribution; Then, it is projected onto the tangent space at $z_0$ using the projection matrix $I_{k-1} - Z_0$, and normalized to unit length, where $Z_0=z_0z_0^\ast$.

We now introduce the estimators considered in this study. First, we apply the Frobenius median to the synthetic data, as introduced in Proposition 1 (iii). Specifically, we compute the sample Frobenius median $\hat{A}_c = \argmin_{A \in \A} \sum_{j=1}^n \|X_j - A\|_F$ and its projection onto $\mathcal{CP}^{k-2}$, that is, $\hat{M}_c = \pi(\hat{A}_c;\mathcal{CP}^{k-2})=\hat{u}_1\hat{u}_1^{\ast}$, where $\hat{u}_1\in \mathcal{CS}^{k-2}$ is a representative pre-shape. 
For ease of presentation, we then perform the Procrustes alignment by rotating $\hat{u}_1$ to minimize its distance to the population pre-shape $z_0$ \citep[p.175]{dryden2016statistical} and obtain $\hat{z}_{\text{EMedian}} = \hat{u}_1 e^{-i \arg(z_0^{\ast} \hat{u}_1)}
$, where $\arg(\cdot)$ is the argument of a complex number. The estimated configuration (with rotation aligned with $z_0$) can be recovered by $H^\top \hat{z}_{\text{EMedian}}$.

We also consider intrinsic alternatives, including the Fr\'echet mean ("IMean") and the Fr\'echet median ("IMedian"), defined respectively as
\begin{align*}
\hat{z}_{\text{IMean}} = \argmin_{z\in \mathcal{CS}^{k-2} } \sum_{j=1}^n \arccos^2(| x_j^* z|) \text{ and } 
\hat{z}_{\text{IMedian}} = \argmin_{z\in \mathcal{CS}^{k-2}} \sum_{j=1}^n \arccos(| x_j^* z|).
\end{align*}
where the Fr\'echet median, also known as the geometric median, provides an intrinsic robust estimator and has been applied in planar shape analysis \citep{fletcher2008robust}. As another robust intrinsic alternative, we also consider the median-of-means ("MoM") approach proposed by \citet{lin2024robust}. Specifically, we randomly split the dataset into seven subsets, compute the Fr\'echet mean within each subset, and take the Fr\'echet median of the resulting Fr\'echet means as the final estimate $\hat{z}_{\text{MoM}}$. 

For all estimators, we use initial values that are orthogonal to $z_0$
to examine their convergence to the global optimum. The entire procedure is repeated for 500 Monte Carlo runs. For each replicate, the estimation errors are measured via the angular distance, $\arccos(|z_0^*\hat{z}|)$. 
Summary statistics of the estimation errors are presented in Table~\ref{tab:simu1_error} and the estimated configurations (with rotation aligned) are presented in Figure~\ref{fig:simu1_shape}.

\begin{table}[ht]
\centering
\resizebox{0.85\textwidth}{!}{
\begin{tabular}{l ccc ccc ccc}
\toprule
 & \multicolumn{3}{c}{Shape 1 ($k=4$)} & \multicolumn{3}{c}{Shape 2 ($k=6$) } & \multicolumn{3}{c}{Shape 3 ($k=13$)} \\
number of outliers & 20 & 40 & 90 & 20 & 40 & 90 & 20 & 40 & 90 \\
\midrule
EMedian  & \textbf{0.0084} & \textbf{0.0091} & \textbf{0.0111} & \textbf{0.0122} & \textbf{0.0129} & \textbf{0.0165} & \textbf{0.0181} & \textbf{0.0189} & \textbf{0.0234} \\
           & (0.003) & (0.003) & (0.004) & (0.003) & (0.003) & (0.004) & (0.003) & (0.003) & (0.004) \\
IMean & 0.1286 & 0.2535 & 0.6047 & 0.1058 & 0.2051 & 0.5109 & 0.0846 & 0.1565 & 0.3976 \\
      & (0.064) & (0.017) & (0.008) & (0.005) & (0.006) & (0.008) & (0.003) & (0.016) & (0.007) \\
IMedian & 0.0182 & 0.0308 & 0.0911 & 0.0216 & 0.0338 & 0.0894 & 0.0263 & 0.0375 & 0.0862 \\
        & (0.109) & (0.070) & (0.041) & (0.116) & (0.057) & (0.004) & (0.051) & (0.003) & (0.005) \\
MoM & 0.0576 & 0.1242 & 0.3699 & 0.0500 & 0.0982 & 0.2693 & 0.0490 & 0.0849 & 0.2074 \\
    & (0.022) & (0.037) & (0.097) & (0.014) & (0.024) & (0.061) & (0.008) & (0.013) & (0.044) \\
\bottomrule
\end{tabular}
}
\caption{Sample medians of estimation errors (i.e., $\arccos(|z_0^*\hat{z}|)$) for planar shape data with standard errors in parentheses (500 replicates).}
\label{tab:simu1_error}
\end{table}
 
The results indicate that EMedian consistently outperforms the intrinsic estimators, with only a slight rise in estimation error as the number of outliers increases.
Although the intrinsic median demonstrates robustness, its estimation error substantially rises as the proportion of contaminated samples grows. The median-of-means estimator only performs well under mild contamination (e.g., 20 outliers) but quickly deteriorates (e.g. 40 outliers) and generally has higher estimation error than IMedian across all contamination levels. 

\begin{figure}[ht]
\centering
\includegraphics[width=0.4\textwidth]{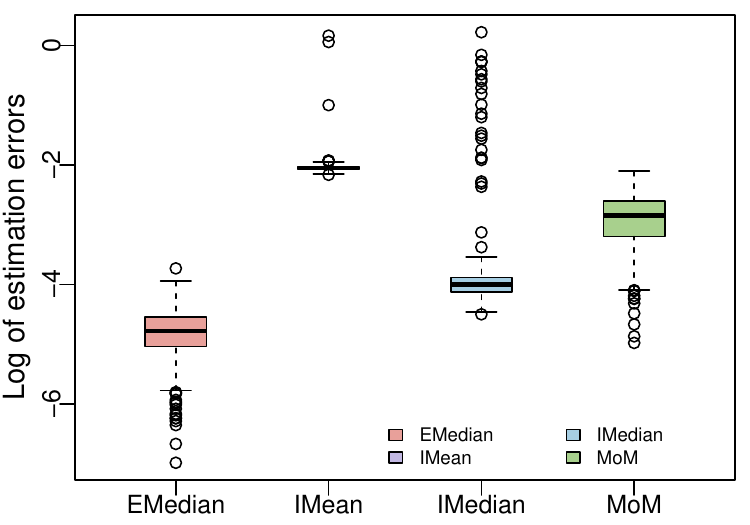}
\caption{Boxplots of the logarithm of estimation errors for planar shape data (Shape 1, 20 outliers, 500 replicates).}
\label{fig:simu1_boxplot} 
\end{figure}

Moreover, the intrinsic median generally exhibits larger standard deviations in estimation errors. Figure~\ref{fig:simu1_boxplot} visually illustrates the variation in the logarithm of estimation errors for each estimator. Evidently, several estimates from IMedian appear beyond the upper whisker, indicating its tendency to converge to a local minimum.

\subsection{Simulation on projective Stiefel manifolds}
\label{sec:simu2}
This section provides numerical studies for the projective Stiefel manifolds, introduced in Section 2.4. Let $X_j = [\pm x_{j1}, \pm x_{j2}, \pm x_{j3}]\in \mathcal{PV}_{3,3}$ for $j=1,\ldots, n$ denote an i.i.d. sample generated from the frame Watson distribution \citep{arnold2013statistics}, with a population mode $M_0=[\pm m_1, \pm m_2, \pm m_3]\in \mathcal{PV}_{3,3}$. 
The probability density function is proportional to $\exp \{\sum_{j=1}^3 \kappa_j (x_j^\top m_j)^2\}$, where $\kappa_j$ denotes the concentration level for each orthogonal axial frame.
In this study, we set the population mode to the canonical basis vectors, i.e., $m_1 = e_1$, $m_2 = e_2$ and $m_3 = e_3$. 
Outliers are generated by rotating $M_0$ along the geodesic path from $m_1$ to the positive orthant (see also \cite{amaral2007pivotal}), leading to a outlier frame $\tilde{M}=[\pm \tilde{m}_1, \pm \tilde{m}_2, \pm \tilde{m}_3]$, where $\tilde{m}_1 = (1,1,1)^\top/\sqrt{3}$, $\tilde{m}_2 = (-2\sqrt{3},\sqrt{3}+3,\sqrt{3}-3)^\top/6$ and $\tilde{m}_3 = (-2\sqrt{3}, \sqrt{3}-3,\sqrt{3}+3)^\top/6$. To contaminate the sample, we randomly replace the sample of size $n=50$ with $n_1 = 0, 5, 15, 20$ observations with $[\tilde{M}]$, corresponding to $0\%, 10\%, 30\%, 40\%$ contamination. 

Furthermore, we consider three concentration settings. For $\kappa=(5,5,5)$, the distribution is rotationally symmetric and moderately concentrated around $M_0$; When $\kappa=(25,5,5)$, it is moderately centered around $\pm m_1$ but highly elliptic/eccentric around $\pm m_2$ and $\pm m_3$; For $\kappa=(50,25,5)$, the concentration around $\pm m_1$ is higher among the three cases, while the remaining frames, $\pm m_2$ and $\pm m_3$, are moderately elliptic/eccentric.
The probability contours of each frame are presented in Figure~S1 of the Supplementary Material. 

We now proceed to the estimation of $M_0$ in the presence of outliers. First, we consider the estimator proposed by \citet{arnold2013statistics}, which is defined as
\begin{equation}\label{e:quotient_mean}
\hat{M}_{\text{mean}} = \argmax_{[U]\in \mathcal{PV}_{3,3}} \sum_{r=1}^3 u_r^\top \bar{T}_r u_r \hskip 0.2truein \hbox{and} \hskip 0.2truein \bar{T}_r = n^{-1}\sum_{j=1}^n \left(x_{jr} x_{jr}^\top - I_3/3\right),
\end{equation}
with $[U]=[\pm u_1, \pm u_2, \pm u_3]$. Note that this estimator is not robust to contamination and is therefore included as a benchmark. 
Robust estimation of $M_0$ can be achieved using the proposed method described in Section~3.5. Specifically, the sample estimator in the ambient space is given as 
\[\hat{B} = \argmin_{B=(B_1,B_2,B_3)\in\mathcal{B}_0} n^{-1}\sum_{j=1}^n\left\{\sum_{r=1}^3 \|B_r - x_{jr}x_{jr}^\top\|_F^2\right\}^{1/2}.\]
Let $(\hat{q}_{11},\hat{q}_{21},\hat{q}_{31})$ be the leading eigenvector for each $\hat{B}_r$ and denote $\hat{Q}_{\epsilon}=[\epsilon_1 \hat{q}_{11},\epsilon_2 \hat{q}_{21},\epsilon_3 \hat{q}_{31}]$. The sample median is given by 
\begin{equation}\label{e:quotient_median}
\hat{M}_{\text{median}}=\pi(\hat{Q}_{\epsilon};\mathcal{V}_{3,3})  
\end{equation}
with the projection $\pi(Q;\mathcal{V}_{3,3})$ defined in~\eqref{Q_epsilon_rep}. The implementation is provided in Section~S2 of the Supplementary Material. 

To evaluate finite-sample performance, we use the angular distance to measure the estimation error for each axis, that is, $\arccos(|m_j^\top \hat{m}_j|)$, where $m_j,\hat{m}_j\in\mathcal{S}^2$ denote the population frame and the corresponding estimates, respectively, for $j=1,2,3$.
Table~\ref{tab:simu2_error} summarizes the average estimation errors for each axis (see Section~S3.1 of the Supplementary Material for complementary results when $n=200$). 

\begin{table}[htbp]
\centering
\resizebox{0.85\textwidth}{!}{
\begin{tabular}{lccccccccc}
\toprule
& \multicolumn{3}{c}{Case 1: $\kappa=(5,5,5)$} & \multicolumn{3}{c}{Case 2: $\kappa=(25,5,5)$} & \multicolumn{3}{c}{Case 3: $\kappa=(50,25,5)$} \\
\cmidrule(lr){2-4} \cmidrule(lr){5-7} \cmidrule(lr){8-10}
Estimator & $\pm m_1$ & $\pm m_2$ & $\pm m_3$ & $\pm m_1$ & $\pm m_2$ & $\pm m_3$ & $\pm m_1$ & $\pm m_2$ & $\pm m_3$ \\
\midrule
\multicolumn{10}{l}{\textbf{Outlier = 0}} \\
$\hat{M}_{\text{mean}}$     & \textbf{0.0090}& \textbf{0.0090} & \textbf{0.0092} & \textbf{0.0030} & \textbf{0.0065} & \textbf{0.0065} & \textbf{0.0014} & \textbf{0.0022} & \textbf{0.0024} \\
              & (0.0048) & (0.0046) & (0.0046) & (0.0016) & (0.0040) & (0.0040) & (0.0008) & (0.0013) & (0.0013) \\
$\hat{M}_{\text{median}}$   & 0.0097 & 0.0097 & 0.0100 & 0.0034 & 0.0072 & 0.0073 & 0.0016 & 0.0024 & 0.0027 \\
              & (0.0052) & (0.0050) & (0.0051) & (0.0018) & (0.0045) & (0.0045) & (0.0009) & (0.0015) & (0.0015) \\
\midrule
\multicolumn{10}{l}{\textbf{Outlier = 5}} \\
$\hat{M}_{\text{mean}}$     & 0.0633 & 0.0451 & 0.0450 & 0.0626 & 0.0448 & 0.0448 & 0.0625 & 0.0443 & 0.0442 \\
              & (0.0074) & (0.0073) & (0.0071) & (0.0025) & (0.0025) & (0.0025) & (0.0012) & (0.0010) & (0.0013) \\
$\hat{M}_{\text{median}}$   & \textbf{0.0119} & \textbf{0.0112} & \textbf{0.0110} & \textbf{0.0046} & \textbf{0.0080} & \textbf{0.0079} & \textbf{0.0020} & \textbf{0.0028} & \textbf{0.0029} \\
              & (0.0061) & (0.0057) & (0.0058) & (0.0023) & (0.0046) & (0.0047) & (0.0010) & (0.0015) & (0.0016) \\
\midrule
\multicolumn{10}{l}{\textbf{Outlier = 15}} \\
$\hat{M}_{\text{mean}}$     & 0.2387 & 0.1684 & 0.1689 & 0.2374 & 0.1677 & 0.1678 & 0.2370 & 0.1674 & 0.1674 \\
              & (0.0077) & (0.0072) & (0.0073) & (0.0026) & (0.0024) & (0.0024) & (0.0012) & (0.0010) & (0.0013) \\
$\hat{M}_{\text{median}}$   & \textbf{0.0314} & \textbf{0.0229} & \textbf{0.0233} & \textbf{0.0145} & \textbf{0.0129} & \textbf{0.0130} &\textbf{ 0.0059} & \textbf{0.0047} & \textbf{0.0051} \\
              & (0.0096) & (0.0087) & (0.0088) & (0.0033) & (0.0044) & (0.0043) & (0.0015) & (0.0015) & (0.0018) \\
\midrule
\multicolumn{10}{l}{\textbf{Outlier = 20}} \\
$\hat{M}_{\text{mean}}$     & 0.3535 & 0.2494 & 0.2493 & 0.3519 & 0.2483 & 0.2482 & 0.3514 & 0.2479 & 0.2479 \\
              & (0.0067) & (0.0063) & (0.0062) & (0.0022) & (0.0021) & (0.0021) & (0.0011) & (0.0008) & (0.0011) \\
$\hat{M}_{\text{median}}$   & \textbf{0.0575} & \textbf{0.0412} & \textbf{0.0412} & \textbf{0.0287} & \textbf{0.0221} & \textbf{0.0221} & \textbf{0.0110 }& \textbf{0.0081} & \textbf{0.0085} \\
              & (0.0108) & (0.0099) & (0.0097) & (0.0042) & (0.0042) & (0.0041) & (0.0017) & (0.0015) & (0.0018) \\
\bottomrule
\end{tabular}
}
\caption{Average estimation errors for each axis (1000 replicates, $n=50$). Standard deviations are shown in parentheses.}
\label{tab:simu2_error}
\end{table}

The results indicate that both estimators produce accurate estimates when there are no outliers, where $\hat{M}_{\text{median}}$ achieves slightly larger but negligible errors. Furthermore, ellipse-like structures in the distribution (case 2 and case 3) have little effect on the estimation accuracy.
In the presence of outliers, as anticipated, the estimation errors of $\hat{M}_{\text{mean}}$ deteriorate rapidly, while those of $\hat{M}_{\text{median}}$ remain small and stable, even with 20 outliers (40\% contamination). This demonstrates the strong robustness of our method.

\begin{figure}[ht]
\centering
\begin{subfigure}[b]{0.25\linewidth}
    \centering
    \includegraphics[width=\linewidth]{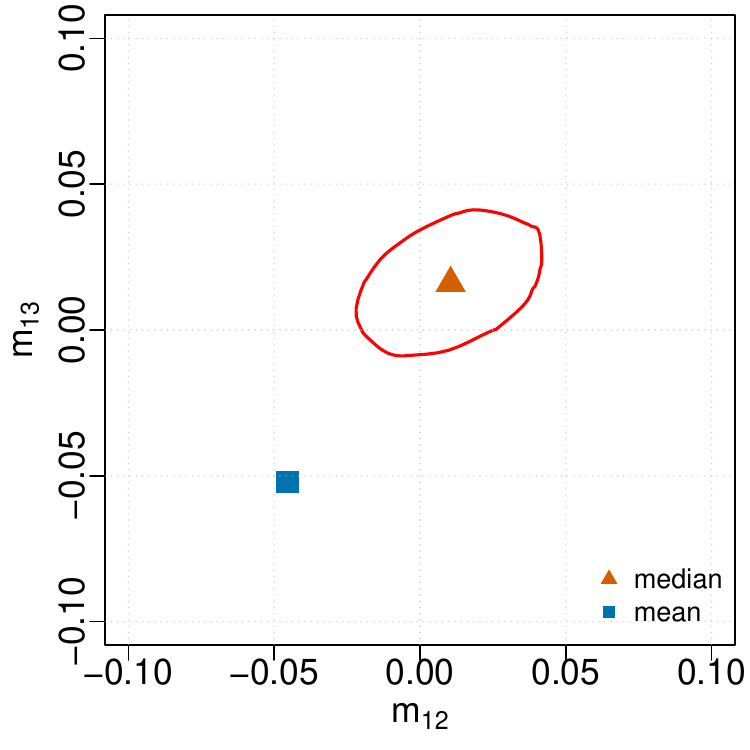}
\end{subfigure}
\hfill
\begin{subfigure}[b]{0.25\linewidth}
    \centering
    \includegraphics[width=\linewidth]{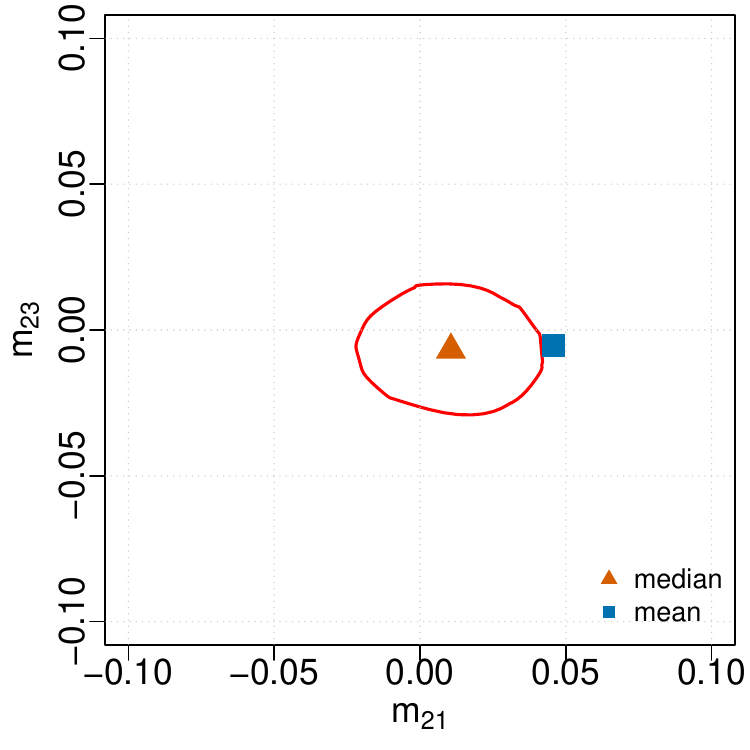}
\end{subfigure}
\hfill
\begin{subfigure}[b]{0.25\linewidth}
    \centering
    \includegraphics[width=\linewidth]{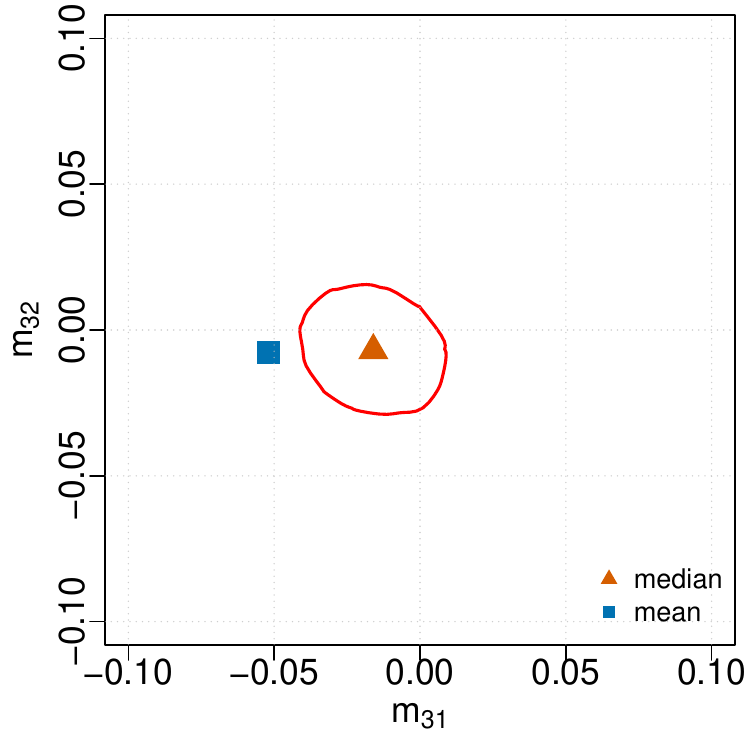}
\end{subfigure}
\caption{The estimates of $\hat{M}_{\text{mean}}$ and $\hat{M}_{\text{median}}$ for Case 1 with $\kappa=(5,5,5), n=50$ and 5 outliers, where the red circles are the 95\% pivotal bootstrap confidence ellipses of $\hat{M}_{\text{median}}$. The population axial frame lies at the origin.}
\label{fig:simu2_est}
\end{figure}

To visually illustrate the influence of the outliers, we provide the estimates of $\hat{M}_{\text{mean}}$ and $\hat{M}_{\text{median}}$ for Case 1 with $\kappa=(5,5,5)$ and 5 outliers in Figure~\ref{fig:simu2_est}. The pivotal bootstrap confidence regions are also attached. Details on the pivotal bootstrap and the construction of the confidence ellipse are provided in Section~S2 of the Supplementary Material.

Figure~\ref{fig:simu2_est} shows that, under high concentration, even a small proportion of outliers (10\% in this case) noticeably shifts $\hat{M}_{\text{mean}}$ away from the population frames. In contrast, $\hat{M}_{\text{median}}$ remains close to the origin, with its confidence ellipse containing the population frames. This observation highlights the robustness of our method.

\subsection{Earthquake moment tensor data}\label{sub:earthquake}
A seismic moment tensor is a $3\times 3$ symmetric matrix, typically constrained to have zero trace, for earthquake faulting events. It encodes the earthquake information carried by the seismic waves, including the orientation of the fault plane, earthquake magnitude and radiation patterns. 
To help describe the orientation of the fault plane and the direction of slip, seismologists often examine the eigenvectors of the moment tensor. Ordered by descending eigenvalues, the associated eigenvectors define an orthogonal axial triple $(\pm t,\pm b,\pm p)\in \mathcal{PV}_{3,3}$, where $\pm  t$ corresponds to the T axis (the axis of tensional stress), $\pm  b$ to the B axis (the null axis) and $\pm p$ to the P axis (the axis of compressional stress). A comprehensive geophysical introduction can be found in \citet{stein2009introduction}, while \citet{arnold2013statistics} provide a detailed statistical analysis for such data.

In this section, we apply our method to estimate the orthogonal axial frames for earthquake events in Papua New Guinea and the Solomon Islands from 2006 to 2016. 
We analyse the moment tensors reported in Table 1 of \citet{hejrani2017centroid} and subdivide the study area into eight regions, following their subdivisions: regions 2 to 4 covering the Bismarck Sea Seismic Lineation as well as the New Guinea Trench in region 1, whereas regions 5 to 8 cover the Papua New Guinea and Solomon islands subduction zone. Figure~S3 in the Supplementary Material displays the map of these earthquakes.

\begin{figure}[ht]
\centering
\begin{subfigure}[b]{0.25\linewidth}
    \centering
    \includegraphics[width=\linewidth]{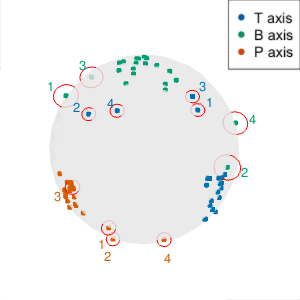}
\end{subfigure}
\hfill
\begin{subfigure}[b]{0.25\linewidth}
    \centering
    \includegraphics[width=\linewidth]{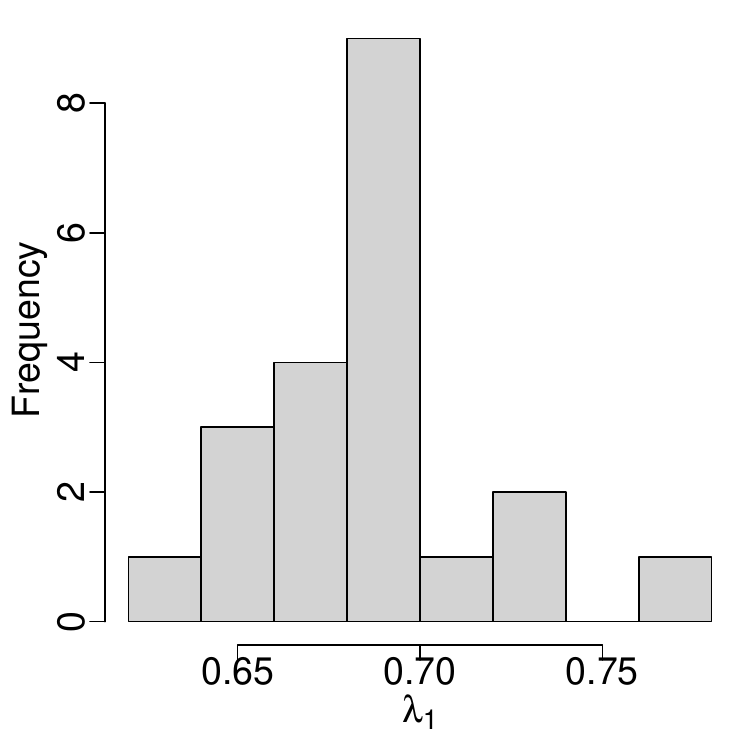}
\end{subfigure}
\hfill
\begin{subfigure}[b]{0.25\linewidth}
    \centering
    \includegraphics[width=\linewidth]{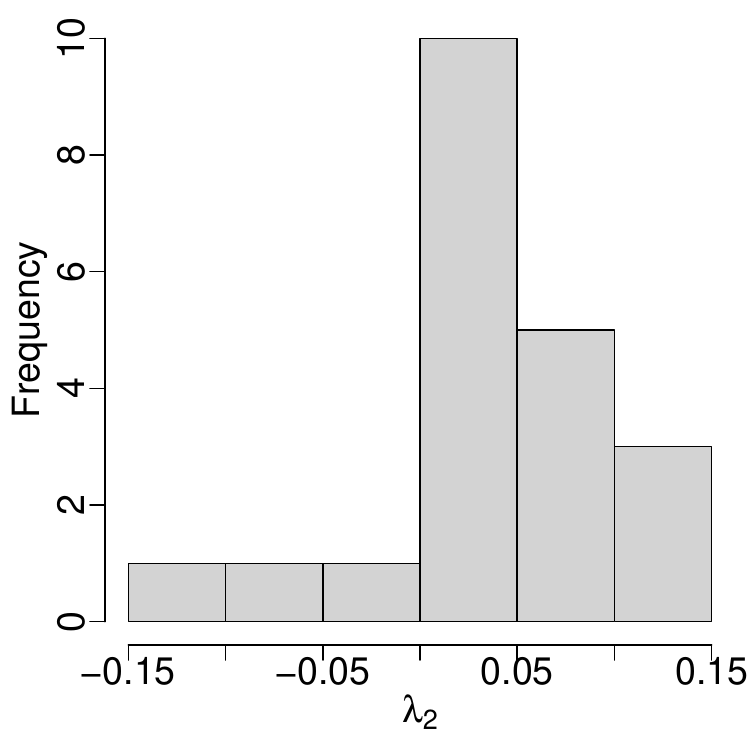}
\end{subfigure}
\caption{The orthogonal axial frames of moment tensors in region 2 (21 events). Left: axes of the moment tensors (up to sign, suspected outliers are circled in red and labelled in numbers). Middle: the largest eigenvalues $\lambda_1$ of the moment tensors. Right: the second largest eigenvalues $\lambda_2$ of the moment tensors.}
\label{fig:papua_axes}
\end{figure}

We focus on events in region 2, where Figure~\ref{fig:papua_axes} displays the axes and the first two eigenvalues of the moment tensors. To ensure that each axis appears only once in the scatter plot, the axes are shifted to the same hemisphere. This plot indicates that the axes are well clustered with four suspected outliers, circled in red.

\begin{figure}[ht]
\centering
\begin{subfigure}[b]{0.25\linewidth}
    \centering
    \includegraphics[width=\linewidth]{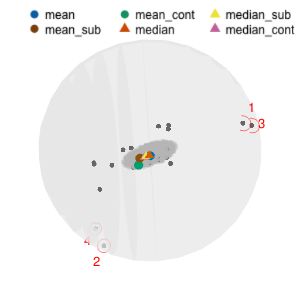}
\end{subfigure}
\hfill
\begin{subfigure}[b]{0.25\linewidth}
    \centering
    \includegraphics[width=\linewidth]{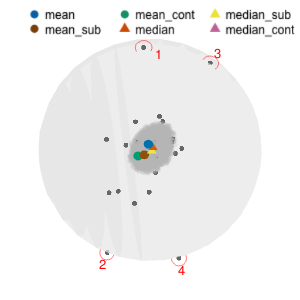}
\end{subfigure}
\hfill
\begin{subfigure}[b]{0.25\linewidth}
    \centering
    \includegraphics[width=\linewidth]{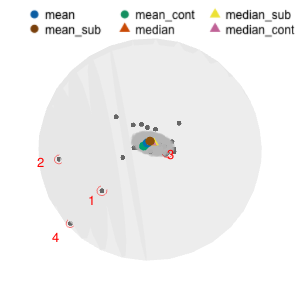}    
\end{subfigure}\\
\begin{subfigure}[b]{0.25\linewidth}
    \centering
    \includegraphics[width=\linewidth]{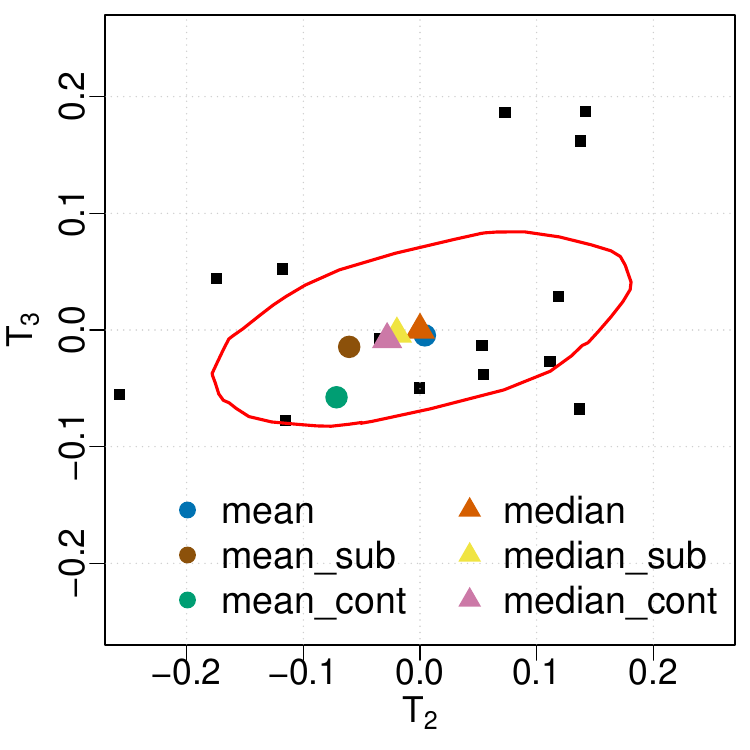}
\end{subfigure}
\hfill
\begin{subfigure}[b]{0.25\linewidth}
    \centering
    \includegraphics[width=\linewidth]{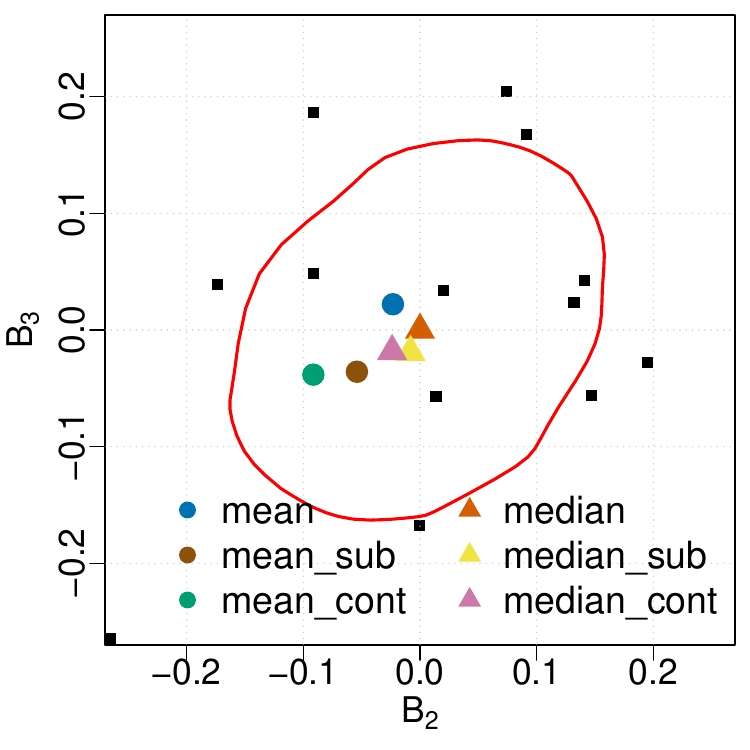}
\end{subfigure}
\hfill
\begin{subfigure}[b]{0.25\linewidth}
    \centering
    \includegraphics[width=\linewidth]{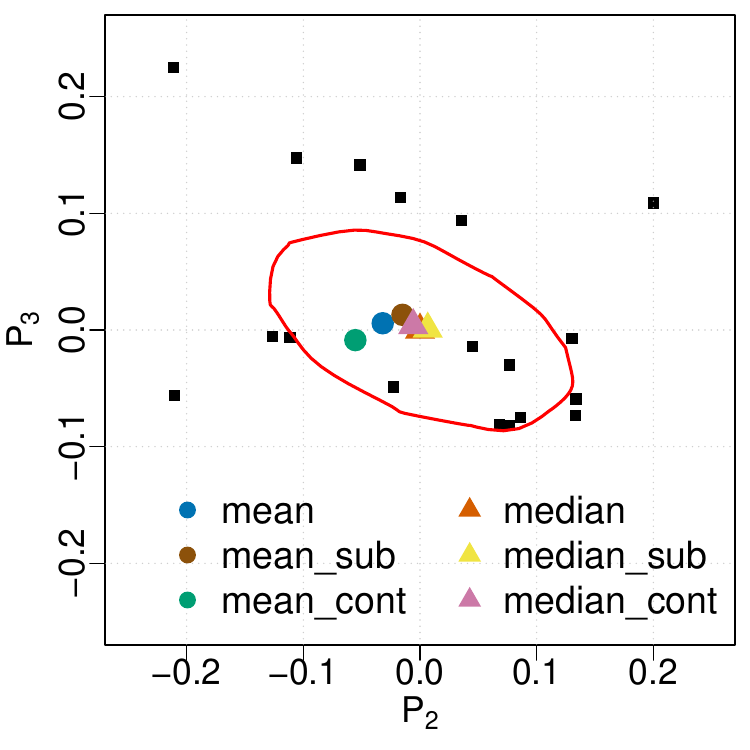}
\end{subfigure}\\
\begin{subfigure}[b]{0.25\linewidth}
    \centering
    \small (a) T axis
\end{subfigure} 
\hfill
\begin{subfigure}[b]{0.25\linewidth}
    \centering
    \small (b) B axis
\end{subfigure}
\hfill 
\begin{subfigure}[b]{0.25\linewidth}
    \centering
    \small (c) P axis
\end{subfigure}
\caption{Scatter plots (top row) and magnified aerial views (bottom row) of the T, B and P axes of moment tensors in region 2. Outliers in region 2 are highlighted with red circles.
The shaded areas (top row) and red circles (bottom row) represent 95\% confidence regions of the spatial median.}
\label{fig:papua_est}
\end{figure}

To estimate the axial frame, we use both the sample mean $\hat{M}_{\text{mean}}$ and the spatial median $\hat{M}_{\text{median}}$, defined in~\eqref{e:quotient_mean} and \eqref{e:quotient_median}, respectively. The first and third rows of Table~\ref{tab:papua_est} report the estimated axes, rounded to three decimal places. The estimates of all axes exhibit negligible differences for both methods, as visually confirmed in Figure~\ref{fig:papua_est}. 
This is likely due to the approximate rotational symmetry of suspected outliers and the moderate concentration of the dataset, both of which mitigate the influence of outliers. To elaborate, we consider two modified datasets: one obtained with two suspected outliers removed (labeled 1 and 3 in Figure~\ref{fig:papua_est}) to break the approximate symmetry (estimators denoted by the postfix "sub"), and another obtained by duplicating the suspected outliers labeled 2 and 4 twice, while removing outliers 1 and 3, to amplify contamination and avoid the approximate symmetry (estimators denoted by the postfix "cont"). 

The bootstrap standard errors for each estimator are computed as follows. Let $\hat{m}_j$ and $\hat{m}_j^{(b)}$ be the estimates of the $j$-th axis and its $b$-th bootstrap estimate, respectively. The estimation error for each axis is measured by the angular distance, i.e., $\arccos(|\hat{m}_j^\top \hat{m}_j^{(b)}|)$ for $b=1,\ldots, 1000$, and the corresponding standard error is taken as the sample standard deviation of these errors.
Moreover, we establish pivotal bootstrap confidence regions for the spatial median, following the procedure as in Section S2 of the Supplementary Material. Additional numerical results supporting the use of the bootstrap confidence region are presented in Section~S3.4. These elliptical confidence regions are plotted in Figure~\ref{fig:papua_est}.  

\begin{table}[ht]
\centering
\resizebox{0.8\textwidth}{!}{
\begin{tabular}{lccc}
\toprule
\textbf{Estimator} & \textbf{T axis} & \textbf{B axis} & \textbf{P axis} \\
\midrule
mean   & $(0.382, -0.472, -0.794)^\top$ & $(0.723, 0.688, -0.061)^\top$ & $(0.575, -0.551, 0.604)^\top$ \\
SE of mean   & 0.0477 & 0.0479 & 0.0236 \\
median &  $(0.384, -0.477, -0.791)^\top$ & $(0.705, 0.704, -0.082)^\top$ & $(0.596, -0.526, 0.607)^\top$ \\
SE of median &  0.0371&  0.0380 & 0.0235\\
mean\_sub & $(0.343, -0.523, -0.781)^\top$ & $(0.739, 0.663, -0.120)^\top$ & $(0.580, -0.536, 0.614)^\top$ \\
SE of mean\_sub & 0.0340 & 0.0341 & 0.0224 \\
median\_sub & $(0.372, -0.492, -0.787)^\top$ & $(0.709, 0.698, -0.101)^\top$ & $(0.599, -0.520, 0.609)^\top$\\
SE of median\_sub & 0.0360 & 0.0366 & 0.0235\\
mean\_cont & $(0.303, -0.519, -0.799)^\top$ & $(0.763, 0.635, -0.123)^\top$ & $(0.571, -0.573, 0.588)^\top$\\
SE of mean\_cont & 0.1220 & 0.1073 & 0.0564 \\
median\_cont & $(0.364, -0.498, -0.787)^\top$ & $(0.720, 0.686, -0.101)^\top$ & $(0.591, -0.530, 0.609)^\top$\\
SE of median\_cont & 0.1195 & 0.1019 & 0.0619 \\
\bottomrule
\end{tabular}
}
\caption{The estimated T, B, and P axes (up to sign) for the moment tensors in region 2, rounded to 3 decimal places.}
\label{tab:papua_est}
\end{table}

When the approximate symmetry of outliers is disrupted, the sample mean shows a more pronounced shift toward the observations labeled 2 and 4 than the sample median (see mean\_sub and median\_sub in Figure~\ref{fig:papua_est}). This shift is further amplified when additional outliers (labeled 1 and 3) are introduced, where the sample mean is pulled toward the outliers and quickly approaches the boundaries of the confidence regions.
In contrast, the spatial median frames remain stable in terms of their shift in Figure~\ref{fig:papua_est}, indicating their robustness against outliers. 

\begin{figure}[htpb]
\centering
\begin{subfigure}[b]{0.2\linewidth}
    \centering
    \includegraphics[width=\linewidth]{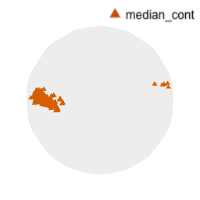}
\end{subfigure}
\hfill
\begin{subfigure}[b]{0.2\linewidth}
    \centering
    \includegraphics[width=\linewidth]{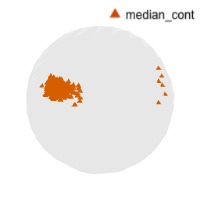}
\end{subfigure}
\hfill
\begin{subfigure}[b]{0.2\linewidth}
    \centering
    \includegraphics[width=\linewidth]{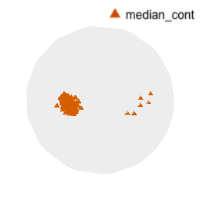}    
\end{subfigure}\\
\begin{subfigure}[b]{0.2\linewidth}
    \centering
    \includegraphics[width=\linewidth]{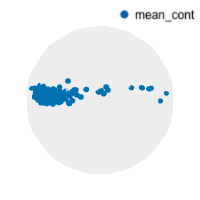}
\end{subfigure}
\hfill
\begin{subfigure}[b]{0.2\linewidth}
    \centering
    \includegraphics[width=\linewidth]{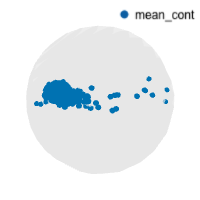}
\end{subfigure}
\hfill
\begin{subfigure}[b]{0.2\linewidth}
    \centering
    \includegraphics[width=\linewidth]{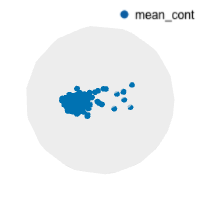}
\end{subfigure}\\
\begin{subfigure}[b]{0.2\linewidth}
    \centering
    \small (a) T axis
\end{subfigure} 
\hfill
\begin{subfigure}[b]{0.2\linewidth}
    \centering
    \small (b) B axis
\end{subfigure}
\hfill 
\begin{subfigure}[b]{0.2\linewidth}
    \centering
    \small (c) P axis
\end{subfigure}
\caption{Bootstrap estimates of the T, B and P axes of the contaminated moment tensors.}
\label{fig:papua_boot}
\end{figure}

Interestingly, the bootstrap standard errors of the sample median in Table~\ref{tab:papua_est} are close to those of the sample mean, and are large particularly for the contaminated dataset. This is likely due to the small sample size (23 earthquakes) and a large proportion of outliers (around 26\%). Therefore, in some bootstrap replications, the resampled data are dominated by outliers, which inflates the estimated standard errors. (See Figure~\ref{fig:papua_boot} for an illustration of the bootstrap estimates.)
All estimates lie in the confidence regions of the spatial median, which appear conservative, likely a consequence of the small sample size and the presence of outliers. In Section~S3.4, we illustrate how sample size affects the confidence region with concentration and contamination level prespecified, which supports our claim.

\section{Conclusion}\label{sec:conc}
We propose a robust and computationally convenient method, the projected Frobenius median (PFM), for location estimation on matrix manifolds. We focus on 4 matrix spaces: Stiefel and Grassmann manifolds, Kendall shape spaces and projective Stiefel manifolds.  Without going into details, we mention other matrix spaces for which the PFM can be used.

Several directions for future study appear particularly promising. First, when handling a compact parameter space, the influence function of the estimator with respect to its dispersion is generally a more appropriate measure of robustness, leading to \emph{SB-robustness} \citep{ko1988robustness}. A careful treatment is required to adapt this concept to matrix manifolds. Second, our application to the earthquake data reveals that orthogonal axial frames often cluster around different axes across regions, motivating the study of the $k$-sample hypothesis testing in projective Stiefel manifolds. While related problems for unsigned directions have been investigated by \citet{amaral2007pivotal}, their extension to the projective Stiefel manifold and possibly in the presence of outliers, remains largely unexplored. 

\section*{Disclosure and Data Availability statements}
The authors report there are no competing interests to declare.
Deidentified data have been made available at the following URL: \url{https://github.com/HourenH/RLocMM}.

\section*{Acknowledgements}
The authors are grateful to the Australian Research Council for supporting this research through grant DP220102232.

\bibliography{Bibliography}

\vfill \eject
\end{document}